\documentclass[useAMS,usenatbib]{mn2e}
\usepackage{mathabx}
\usepackage{longtable}
\usepackage{ulem}

\title[Reduced atomic models for Fe XIII]
{Reduced atomic models for large-scale computations: Fe XIII near-infrared lines} 
\author[Del Zanna, Supriya]{Giulio Del Zanna$^{1,2}$\thanks{E-mail: gd232@cam.ac.uk},
Supriya Hebbur Dayananda$^{3,4}$ \\
$^{1}$ DAMTP, Centre for Mathematical Sciences, University of Cambridge, Wilberforce Road, Cambridge CB3 0WA, UK \\
$^{2}$ School of Physics \& Astronomy, University of Leicester, Leicester  LE1 7RH, UK \\
$^{3}$ Instituto de Astrofísica de Canarias, E-38205, La Laguna, Tenerife, Spain \\
$^{4}$ Departamento de Astrofísica, Facultad de Física, Universidad de La Laguna, Tenerife, Spain
}
\date{Submitted to MNRAS  }

\pagerange{\pageref{firstpage}--\pageref{lastpage}} \pubyear{2024}

\DeclareMathAlphabet{\mathsc}{OT1}{cmr}{m}{sc}
\def\testbx{bx}%
\DeclareRobustCommand{\ion}[2]{%
\relax\ifmmode
\ifx\testbx\f@series
{\mathbf{#1\,\mathsc{#2}}}\else
{\mathrm{#1\,\mathsc{#2}}}\fi
\else\textup{#1\,{\mdseries\textsc{#2}}}%
\fi}

\usepackage{graphicx,threeparttable}
\usepackage{txfonts}

\usepackage{epsfig}
\usepackage{natbib}
\usepackage{color}

\newcommand{\beq}{\begin{equation}}
\newcommand{\eeq}{\end{equation}}

\newcommand{\rsun}{\mbox{\,$\rm R_{\odot}$}}        

\begin{document}

\label{firstpage}
\maketitle

\begin{abstract}
Accurate atomic models for astrophysical plasma
can be very complex, requiring thousands of states.
However, for a variety of applications such as
large-scale forward models of the Stokes parameters of a spectral line in
the solar corona, it is necessary to build much reduced atomic models.
We present two examples of such models, focused on the two
near-infrared \ion{Fe}{xiii} lines observed on the ground at 10750, 10801~\AA.
These lines are primary diagnostics for a range of missions (especially
the Daniel K. Inouye Solar Telescope, DKIST)
to measure electron densities and magnetic fields in the solar corona.
We calculate the Stokes parameters for a range of coronal conditions using
CHIANTI (for intensities) and P-CORONA (for intensities and polarization),
and use P-CORONA and a realistic global  MHD simulation 
to show that the reduced models provide accurate
results, typically to within 5\% those obtained with larger models. Reduced models
provide a significant decrease (over three orders of magnitude) in the computational time
in spectropolarimetric calculations. 
The methods we describe are general and can be applied to a range of
conditions and other ions.  
\end{abstract}


\section{Introduction}

Accurate atomic models for astrophysical plasma
can be very complex, depending on the local environment and
which physical processes need to be included. 
Typically, a large-scale collisional-radiative model needs to be built, to take into account
ionization and recombination via Rydberg states, requiring thousands 
of states \citep[e.g. the helium models by][]{delzanna_etal:2020_he,delzanna_storey:2022}.

On the other hand, computationally intensive calculations are becoming necessary.
For example, to assess the reliability of three-dimensional (3D) MHD models of the extended solar corona 
such as those produced by ``Predictive Science Inc." (PSI\footnote{https://www.predsci.com/portal/home.php})
and compare them to observables \citep[see, e.g.][]{schad_etal:2023} one needs to calculate the local emissivity over a large volume, and then perform integrations along the line-of-sight (ALOS). 
A lot more challenging are 3D spectropolarimetric calculations, discussed below. 
Another obvious example is radiative-transfer calculations in 3D.
In all such cases, it is necessary to build much reduced atomic models, but 
making sure that the results of the computations are `close' to those obtained 
with the full model. 
How close depends on a number of factors. First of all on the accuracy
of the observables. Secondly, on the accuracy of the large-scale atomic models, an issue 
briefly discussed below.

We focus this paper on modelling the \ion{Fe}{xiii} lines at 10750, 10801~\AA\
(vacuum wavelengths)
first discovered by Bernard Lyot in the 1930s \citep[cf.][]{lyot:1939}, as they are the strongest lines in the near infrared (NIR).
There is a growing interest in spectroscopic observations of the solar coronal
forbidden lines in the NIR,
as they offer powerful diagnostics to measure ionization temperatures, electron
densities and chemical abundances, as reviewed in \cite{delzanna_deluca:2018},
and to obtain information on the coronal magnetic field with spectropolarimetry (SP), as described e.g.
in  \cite{judge:1998}, \cite{judge_etal:2013}, and \cite{penn:2014}.

Regular observations of the \ion{Fe}{xiii} NIR lines have been  carried out with the
Coronal Multichannel Polarimeter (CoMP) instrument, described by \cite{tomczyk_etal:2008},
and, together with more lines with its  upgraded version UCoMP.
They are also a target for the 
Coronal Solar Magnetism Observatory (COSMO),
see e.g.  \cite{tomczyk_etal:2016,landi_etal:2016}.

The airborne infrared spectrometer (AIR-Spec) has surveyed the NIR during the
2017 and 2019 eclipses \citep[cf.][]{madsen_etal:2019,Samra2021,samra_etal:2022,delzanna_etal:2023}
as a pathfinder for future observations  with the Daniel K. Inouye Solar Telescope
(DKIST, see \citealt{rimmele_etal:2015}) CryoNIRSP
spectropolarimeter \citep{fehlmann_etal:2016}.
CryoNIRSP is now regularly observing several  NIR lines.

The \ion{Fe}{xiii} NIR lines are commonly used to measure the electron density,
although direct measurements cannot be provided: as their intensities are 
affected by both collisions with free electrons and protons
and  photo-excitation (PE) from the disk radiation, a prior knowledge of the
density ALOS is needed \citep[cf. the appendix in][]{delzanna_etal:2023}.
In conjunction with measurements of EUV lines from the same ion, they provide the opportunity
to estimate the density distribution along the line of sight, and assess
if non-Maxwellian electron distributions are present
\citep{dudik_etal:2014_fe,dudik_etal:2021_eis_comp}.

One long-standing problem in solar physics has been the estimate 
of the local magnetic field.  
With some assumptions on the Alfven waves in the corona, the
\ion{Fe}{xiii} NIR lines can be used to infer the coronal magnetic field, as
described in \cite{yang_etal:2020}.
However, a more direct way to obtain information on the coronal magnetic field
is with spectropolarimetry.

As in the case of the electron density, a direct inversion of observables to obtain the magnetic field is not reliable due to the contribution ALOS in the observed signal. Therefore, it is necessary to synthesize the Stokes parameters using forward modelling and compare them with observations to validate the results and extract information about the magnetic field.
Recently, DKIST CryoNIRSP has obtained  unprecedented Stokes profiles of the stronger 10747~\AA\ line \citep{schad_etal:2024}. 
The forward modelling tool pyCLE, developed by \cite{schad_dima:2020},
was then used to calculate the Stokes $V$ signal using as input an MHD model. The resulting $V$ signal was not close to the observed one. These studies led them to conclude that the longitudinal coronal magnetic field strengths at greater heights above the solar limb may be higher than those predicted by conventional global coronal models. They attribute this difference to limitations in their study, including the relatively low resolution of the employed model and the simplified treatment of the lower boundary conditions.

It has long been known that the populations of the ground configuration,
producing the NIR lines, are strongly affected by PE and proton collisions;
also,  both level and magnetic sublevel populations are affected by cascading from higher levels, populated
by electron collisions \citep[see, e.g.][]{house:1977,Sahal-Brechot:1977,judge_etal:2006}.
Cascading is particularly important as  it increases significantly  the 
populations and as via collisional coupling  destroys  the polarization in the forbidden lines. 
In order to account for the cascading effects 
from the higher states to the ground configuration we need to consider large multi-level  atomic models. 

 This makes the calculations of the intensities (and all the other Stokes profiles)  for 3-D forward models extremely time-consuming and it is 
therefore important to reduce the atomic models.
An approach was developed by \cite{judge_etal:2006}  and
used within the “Coronal Line Emission” (CLE) FORTRAN programs, 
which have been
integrated in the SolarSoft FORWARD  IDL package \citep{gibson_etal:2016}.
The starting point  was  older and incomplete atomic rates for an
ion model comprising of 27 states. Such model was reduced to a 3-state model,
and ad-hoc corrections were included to take into account the effects
of cascading from the higher states (by increasing the collisional rates),
and of the associated depolarization (with an ad-hoc parameter), by comparing the results of
the 3-state with those from the 27-state model.

The need to consider improved atomic rates and
a model larger than the 27-states one to account of the cascading effects  was pointed out by one of us (GDZ) during the
first DKIST coronal workshop at Maui to T. Schad; 
 a Python version of CLE (pyCLE) was developed and used by \cite{schad_dima:2020} to show how the predicted
intensities of the NIR lines change by reducing the  
 CHIANTI version 8 \citep{delzanna_chianti_v8} 749 states model to
the lowest 27 and 100 states. They noted that the largest model
adopted by \cite{judge_etal:2006} also had 27 states, and showed that
differences in the line intensities of the order of 10\% can occur, when compared
to the 749-state model. 
They concluded that at least 100 atomic states were needed in order to 
obtain accurate results while undertaking SP computations.

We note that the magnetic field information in the Fe {\sc xiii} forbidden lines is contained in the Stokes parameter ratios $U/Q$ and $V/I' (I^{\prime}= {\rm d}I / {\rm d}\lambda )$. These lines, being in the saturated Hanle regime for coronal magnetic field strengths, are sensitive only to the magnetic field orientation through Hanle effect, determined using the ratio $U/Q$, and not to the magnetic field strength. The line-of-sight component of the magnetic field strength can be derived from $V/I'$.
Given this, if one considers the Stokes parameter ratios $V/I^{\prime}$ and $U/Q$, 
the effects of the cascades from higher states appear to cancel out, considering their 
expressions \citep{casini_judge:1999}.
 However, we demonstrate in the Appendix that this cancellation occurs primarily for large atomic models.
  For  smaller atomic states, the cascading effects in the ratios do not cancel out.
Moreover, obtaining these ratios requires calculating the full Stokes profiles ($I$, $Q$, $U$, $V$) by solving a multi-level atomic system for the density matrix elements and integrating ALOS to enable meaningful comparison with observations, a process that benefits from the use of reduced atomic models. 
Also,  calculating $V/I^{\prime}$  produces additional noise to already very noisy data, and 
it is probably one of the reasons why the previous studies by 
\cite{lin_etal:2004,schad_etal:2024} considered all the Stokes profiles to estimate the
magnetic field. 

Recently, P-CORONA, a general suite of codes  to perform forward modelling and
predict the SP signal in allowed and forbidden lines 
was developed by one of us (SD), see \cite{2021ApJ...920..140S}, as part of the
POLMAG EU-funded project\footnote{http://research.iac.es/proyecto/polmag}.
The above  three suites of SP codes  are all based on the density matrix
formalism and various approximations described
by \cite{landi_book:2004}, although some differences are present.

Our ultimate goal is to improve the atomic modelling for \ion{Fe}{xiii}, but also benchmark the various codes.
This is carried out 
within an ISSI team\footnote{https://teams.issibern.ch/middlecorona/}.
Such comparisons are important. For example, whilst developing pyCLE,
\cite{schad_dima:2020} found a significant bug in the CLE FORTRAN code
which was subsequently fixed. 

In this paper, we provide some example reduced models, and show how spectral line 
intensities and Stokes parameters differ from those obtained with the full model,
for a range of cases. The methods we propose are general and we plan to extend them to other 
spectral lines in a future paper.

\section{Methods}

The largest-scale scattering calculations for \ion{Fe}{xiii} were
 carried out by 
\cite{delzanna_storey:2012_fe_13} as part of a long-term programme to
improve the atomic data for the soft X-rays.
The main calculations adopted 
 the  $R$-matrix suite of codes and included a complete set
of the main  $n=4$ configurations, producing  749 fine-structure states.
Significant resonance enhancement for some $n=4$ configurations was found. 

A distorted-wave calculation which included the main
$n=5,6$ configurations, giving rise to 3066 fine-structure levels was also
carried out, finding that cascading from the $n=5,6$ states affected the
lower level populations by less than 10\%.  Given the limitations  of the
CHIANTI model atoms at the time, only the model with the 749 states was included in
version 8 \citep{delzanna_chianti_v8}, and is still the same in the latest version
11 \citep{dufresne_etal:2024}.

The obvious question arises: what is the uncertainty in the large atomic models,
to be used as a guide to how close the reduced models should be?
The answer is not simple. One of us (GDZ) developed a simple method whereby different 
atomic calculations (of similar accuracy in principle) are compared, and from the 
scatter of the values an uncertainty for each single rate is obtained. 
These uncertainties are then propagated with a Monte Carlo method, by varying the rates 
randomly. The scatter in the resulting emissivities in a single spectral line 
is a measure (more of an upper limit) of the uncertainty. Such method was applied to the EUV lines of  \ion{Fe}{xiii} 
\citep{yu_etal:2018} and to the NIR lines \citep{yang_etal:2020}. 
The estimated uncertainty in the NIR ratio is about 5\% at a density of 10$^8$ cm$^{-3}$.

Considering also that it would be difficult to obtain measurements of Stokes 
parameters any  better than 5\%, we think that for most applications 
a reduced model which gives results to within about 5\% those of the full model
is a reasonable choice. 

The main physical processes to consider in these computations are collisional excitation and 
de-excitation processes due to electrons, protons, and PE of the disk radiation. 
It has long been known that proton rates affect the populations of the ground configuration 
of coronal ions. We use here the \mbox{CHIANTI} proton rates which are $J$-resolved. 
The same rates are used by  PyCLE and P-CORONA, although we note that the 
$M$-resolved rates should be used instead. The issue of calculating $M$-resolved  proton rates 
for coronal ions is a complex one and will be addressed in a future paper.

 PE from the disk  is simple to include: as the 
radiation is effectively 
a continuum, it is well approximated with a black-body of 6100 K at the wavelengths 
of the   \ion{Fe}{xiii} NIR lines.
  The PE rate when the radiation is a continuum has traditionally been included in
  the literature  as a modification of the A-value.
The  CHIANTI codes follow that approach by first constructing a matrix of  photoexcitation rates and
one for the corresponding stimulated emission rates (which are normally negligible).
These matrices are then added  to the matrix 
of the A-values before solving for the level populations. 
As already mentioned, PE is a strong effect, so it is impossible to infer the 
electron density (average ALOS) without knowing a priori its distribution. 
Hence, forward modelling needs to be carried out. 
The same reasoning applies to the Stokes parameters: one needs to calculate them 
for each point in a coronal volume, then perform integrations ALOS. This is 
easily carried out by P-CORONA, but the computing times for large models are 
very long, as we point out below with some examples. 

\subsection{Selecting  key states}

\begin{table}
\caption{Selected 55 states  for  \ion{Fe}{xiii}.
}
\begin{center}
\begin{tabular}[c]{@{}rlclll@{}}
 \hline\hline \noalign{\smallskip}
 Lev. &  Conf.  & LSJ & $E_{\rm exp}$ &   $E_{\rm t}$   \\
\noalign{\smallskip}\hline\noalign{\smallskip}                                  
1 &   3s$^2$ 3p$^2$ & $^3$P$_{0}$ &  0.000 &  0.000 \\
    2 &   3s$^2$ 3p$^2$ & $^3$P$_{1}$ &  0.085 &  0.081 \\
    3 &   3s$^2$ 3p$^2$ & $^3$P$_{2}$ &  0.169 &  0.167 \\
    4 &   3s$^2$ 3p$^2$ & $^1$D$_{2}$ &  0.438 &  0.451 \\
    5 &   3s$^2$ 3p$^2$ & $^1$S$_{0}$ &  0.834 &  0.858 \\
    6 &       3s 3p$^3$ & $^5$S$_{2}$ &  1.956 &  1.911 \\
    7 &       3s 3p$^3$ & $^3$D$_{1}$ &  2.617 &  2.608 \\
    8 &       3s 3p$^3$ & $^3$D$_{2}$ &  2.619 &  2.610 \\
    9 &       3s 3p$^3$ & $^3$D$_{3}$ &  2.644 &  2.635 \\
   10 &       3s 3p$^3$ & $^3$P$_{0}$ &  2.997 &  3.001 \\
   11 &       3s 3p$^3$ & $^3$P$_{1}$ &  3.004 &  3.008 \\
   12 &       3s 3p$^3$ & $^3$P$_{2}$ &  3.010 &  3.013 \\
   13 &       3s 3p$^3$ & $^1$D$_{2}$ &  3.302 &  3.320 \\
   14 &       3s 3p$^3$ & $^3$S$_{1}$ &  3.786 &  3.839 \\
   15 &    3s$^2$ 3p 3d & $^3$F$_{2}$ &  3.920 &  3.968 \\
   16 &    3s$^2$ 3p 3d & $^3$F$_{3}$ &  3.981 &  4.031 \\
   17 &       3s 3p$^3$ & $^1$P$_{1}$ &  3.992 &  4.049 \\
   18 &    3s$^2$ 3p 3d & $^3$F$_{4}$ &  4.073 &  4.121 \\
   19 &    3s$^2$ 3p 3d & $^3$P$_{2}$ &  4.432 &  4.507 \\
   20 &    3s$^2$ 3p 3d & $^3$P$_{1}$ &  4.510 &  4.580 \\
   21 &    3s$^2$ 3p 3d & $^1$D$_{2}$ &  4.546 &  4.619 \\
   22 &    3s$^2$ 3p 3d & $^3$P$_{0}$ &  4.570 &  4.632 \\
   23 &    3s$^2$ 3p 3d & $^3$D$_{1}$ &  4.616 &  4.687 \\
   24 &    3s$^2$ 3p 3d & $^3$D$_{2}$ &  4.641 &  4.715 \\
   25 &    3s$^2$ 3p 3d & $^3$D$_{3}$ &  4.640 &  4.719 \\
   26 &    3s$^2$ 3p 3d & $^1$F$_{3}$ &  5.075 &  5.180 \\
   27 &    3s$^2$ 3p 3d & $^1$P$_{1}$ &  5.201 &  5.305 \\
   28 &          3p$^4$ & $^3$P$_{2}$ &  5.389 &  5.492 \\

   31 &          3p$^4$ & $^1$D$_{2}$ &    -   &  5.715  \\

   33 &    3s 3p$^2$ 3d & $^5$F$_{2}$ &    -   &  5.797  \\
   34 &    3s 3p$^2$ 3d & $^5$F$_{3}$ &    -   &  5.829  \\
   35 &    3s 3p$^2$ 3d & $^5$F$_{4}$ &    -   &  5.875  \\

   42 &    3s 3p$^2$ 3d & $^3$F$_{2}$ &  6.102 &  6.217 \\

   45 &    3s 3p$^2$ 3d & $^5$P$_{3}$ &   -    &  6.430   \\

   49 &    3s 3p$^2$ 3d & $^3$P$_{2}$ &   -    &  6.608   \\

   52 &    3s 3p$^2$ 3d & $^3$G$_{3}$ &   -    &  6.908   \\

   56 &    3s 3p$^2$ 3d & $^3$D$_{2}$ &  6.869 &  6.991 \\

   59 &    3s 3p$^2$ 3d & $^1$F$_{3}$ &   -    &  7.110   \\
   60 &    3s 3p$^2$ 3d & $^3$F$_{2}$ &  7.089 &  7.217 \\
   61 &    3s 3p$^2$ 3d & $^3$F$_{3}$ &  7.140 &  7.268 \\

   64 &    3s 3p$^2$ 3d & $^1$P$_{1}$ &   -    &  7.345  \\
   65 &    3s 3p$^2$ 3d & $^3$F$_{4}$ &  7.218 &  7.346 \\

   72 &    3s 3p$^2$ 3d & $^3$D$_{2}$ &  7.447 &  7.580 \\

   74 &    3s 3p$^2$ 3d & $^3$F$_{2}$ &   -    &  7.667  \\
   75 &    3s 3p$^2$ 3d & $^3$F$_{3}$ &   -    &  7.708  \\
   76 &    3s 3p$^2$ 3d & $^3$F$_{4}$ &   -    &  7.731  \\
   77 &    3s 3p$^2$ 3d & $^1$D$_{2}$ &   -    &  7.765  \\

   81 &    3s 3p$^2$ 3d & $^1$D$_{2}$ &   -    &  8.045  \\

   83 &    3s 3p$^2$ 3d & $^1$P$_{1}$ &   -    &  8.260  \\
   84 &    3s 3p$^2$ 3d & $^3$P$_{2}$ &  8.179 &  8.311 \\
   85 &    3s 3p$^2$ 3d & $^3$P$_{1}$ &   -    &  8.379  \\
   86 &    3s 3p$^2$ 3d & $^3$P$_{0}$ &   -    &  8.381   \\
   87 &    3s 3p$^2$ 3d & $^1$F$_{3}$ &   -    &  8.602   \\

   90 &   3s$^2$ 3d$^2$ & $^3$F$_{4}$ &   -    &  8.867   \\
   91 &    3s 3p$^2$ 3d & $^1$D$_{2}$ &   -    &  8.897  \\
\noalign{\smallskip}\hline\noalign{\smallskip}                                   
\multicolumn{5}{l}{\small $E_{\rm exp}$ are the experimental energies in Rydbergs.}\\
\multicolumn{5}{l}{\small $E_{\rm t}$ are those obtained from the scattering target,}\\
\multicolumn{5}{l}{\small see \cite{delzanna_storey:2012_fe_13}. Lev. is the level } \\
\multicolumn{5}{l}{\small number of the CHIANTI 749-states model.} \\
\end{tabular}
\normalsize
\end{center}
\normalsize
\label{tab:energies1}
\end{table}

The method of finding a reduced model is very simple. We calculate the populations
of the ground configuration states for a range of parameters, including proton rates 
and PE, then trace which higher states contribute via cascading, 
and include in the model only 
those above some threshold (a fraction of a percent). 
For example,  if we consider the 10798~\AA\ line we need to 
study in detail the 
population of the upper state, the  3s$^2$ 3p$^2$ $^3$P$_2$. 
At 1.1\rsun, log Ne=8, and with a 6100 K black-body
the population of this state is by 67\% due to cascading.

The largest contributor to cascading (14\%) comes from level 14 (3s 3p$^3$ $^3$S$_1$), which is populated by
cascading only by 3\%, with level 84 (3s 3p$^2$ 3d $^3$P$_2$) being the main contributor,
which we have included.
13\% of the population comes from the next higher level 4 (3s$^2$ 3p$^2$ $^1$D$_2$) within the ground
   configuration, which in turn is
populated by 61\% via cascading, mostly from many states of the 3s 3p$^3$
and 3s$^2$ 3p 3d configurations, which in turn are also  populated partly by cascading
which needs to be followed.
The next state contributing is level 20 (3s$^2$ 3p 3d $^3$P$_1$) which produces the resonance
transition, hence is  almost entirely populated by  excitation from the ground state.
The next contribution, by 5\%, is from level 9 (3s 3p3 $^3$D$_3$), which in turn is populated
   for 15\% by cascading from level 28 (3p$^4$ $^3$P$_2$).
Another 5\% comes from level 25   (3s$^2$ 3p 3d $^3$D$_{3}$)
   which has a small cascading contribution (9\%) with over 3\% from states 84 and 90, which we have included in the model.
   Levels 11, 16, 23, and 24 each contribute about 3\%, so they need to be included as well
as any higher states with significant cascading contributions to those levels.
Finally, about 5\% comes from the  6,15, and 19 levels.

The same procedure was applied to the 3s$^2$ 3p$^2$ $^3$P$_1$. 
 This is to make sure that all the main states contributing to its population
by cascading effects are included.

It is clear that the number of higher states that need to be included initially grows 
considerably. However, contributions gradually decrease.
Typically, higher states are populated by electron collisions from the ground 
configuration, with increasingly smaller cascading contributions for increased excitation energy. 

It is therefore possible to select
a reasonably small number of states which produce the main cascading and populations
of the two main states within a few percent. There are clearly many choices depending
on which accuracy is needed. 
 We present a newly constructed model with a {\it selected} set of 55 states. This includes 
all the lowest 28 states and selected 27 states which are the major contributors to the 
populations of the two main states via cascading.
For an accuracy of 5 percent (or better) we suggest 
this model of {\it selected} 55 $J$-resolved states, 
listed in Table\ref{tab:energies1} and provided in CHIANTI format. 
We used this model for line intensity calculations with the CHIANTI programs and 
the polarization signal with  P-CORONA calculations.  
The model gives very good results for a range of coronal conditions, as shown below.

Clearly, the method can easily be applied to other physical conditions and automatized. 
The method focuses on the two NIR transitions. However, we have also checked that 
the model produces populations of the ground state close to those calculated with the 
full model, as shown in the Appendix. 
As almost all the population of the ion for coronal conditions is in the $^3$P states,
these are the states populating the higher ones by electron collision. 
Therefore, the model also produces accurate populations for any higher state where cascading is negligible.

\subsection{Merging states}

The density matrix formalism and approximations described
by \cite{landi_book:2004} requires $J$-resolved states. 
However, if one is interested in modelling just the intensities of the lines, 
a better approach is that of merging states, or super-levels. 
As in the previous case, there are  many options. We have chosen to keep the 
lowest 27 states $J$-resolved, and added six super-levels
(described in Table~\ref{tab:merged}), for a total of 33  states. 
We also provide CHIANTI files to be used. 

The current standard CHIANTI structure unfortunately only allows 
 $LSJ$ states. However, in principle each state could
have any label, the only quantities used from the energy file 
for the calculations are the 
energies of the states (in inverse cm) and the multiplicity, calculated
as $g=2J+1$. When merging the $k$ states we have summed the multiplicities to 
obtain the multiplicity of the merged state, then wrote in the file 
a corresponding fictitious $J$ value.
The energy $E_m$ of the merged state is the weighted average:
$E_m = (\sum_k  g_k E_k ) / \sum_k  g_k$. 
From these averaged values we have recalculated averaged wavelengths for the 
transitions to the lower states. 

We have then calculated the excitation rate coefficients from 
each of the lower 27 states to each of the merged states. CHIANTI uses a 
 \cite{burgess_tully:1992} scaled value of 
the adimensional normalised  rate coefficient called the effective collision 
strength $\Upsilon$, calculated from the cross-section by assuming a 
Maxwellian electron distribution.
We have first obtained 18  $\Upsilon_{ij}$ values over a very large temperature
range, from 10$^4$ to 10$^8$ K, with steps of 0.2 dex around the peak formation
 temperature, then obtained the 
$\Upsilon_{im} = \sum_k \Upsilon_{ik} $. After that we have scaled the 
$\Upsilon_{im}$ and extrapolated the values to obtain the limit points,
to produce the CHIANTI-format rates.
When merging transitions of different  \cite{burgess_tully:1992} type, 
we have chosen to adopt type 2  (forbidden transition), and 
in any case visually inspected each scaled rate.
 We note that the choice of type of transition for the scaling 
  is essentially irrelevant. It is just a transformation
  so the interpolation is performed on more smoothly-varying rates.
  As long as there are sufficient grid points.
  The extrapolation to the limit points is also irrelvant, given the
large temperature range the data are provided. 
There is no need to consider the excitation rates among the merged states
as they are essentially populated from the ground configuration states. 

Regarding the A-values, we have kept those among the lowest 27 states.
The  last step was to calculate the A-values $A_{im}$ between the 
$J$-resolved 27 states $i$ and the merged states $m$.
In order to satisfy  Einstein's relations between the radiative coefficients, they are: 
$A_{im} = (\sum_k  g_k A_{ik}) / \sum_k  g_k$. 
We have neglected radiative transitions among the merged states as 
all the main decays from the higher states are to the lower 27 states,
even though in principle they could be added with the appropriate averaging.

\begin{table}
  \caption{Merged states}
\begin{center}
\begin{tabular}[c]{@{}rllll@{}}
   \hline\hline \noalign{\smallskip}   
Lev & CHIANTI &   Configuration      & states merged \\
    & Levs    &                &   \\
    \noalign{\smallskip}\hline\noalign{\smallskip}
28 & 28--34     & 3p$^4$         & $^3$P, $^1$D$_{2}$  \\
29 & 32--44     &   3s 3p$^2$ 3d & $^5$F, $^5$D,  $^3$F \\
30 & 45, 47--54 &  3s 3p$^2$ 3d  & $^5$P, $^3$P,$^3$G \\
31 & 55--64     & 3s 3p$^2$ 3d   & $^3$D, $^1$G$_{4}$,$^1$F$_{3}$,$^3$F, $^3$D, $^1$P$_{1}$ \\ 
32 & 65--77     & 3s 3p$^2$ 3d   & $^3$F, $^3$D, $^3$P, $^3$S$_{1}$, $^1$D$_{2}$ \\
33 & 78--87,91  &  3s 3p$^2$ 3d  & $^3$D,  $^1$D$_{2}$,$^1$S$_{0}$, $^1$P$_{1}$, $^3$P,$^1$F$_{3}$,$^1$D$_{2}$ \\
  \noalign{\smallskip}\hline\noalign{\smallskip}
\multicolumn{4}{l}{\small  Lev is the merged level number, Levs the original level numbers.} \\
\end{tabular}
\normalsize
\end{center}
\normalsize
\label{tab:merged}
\end{table}

\section{Results}

\subsection{Validation of the reduced atomic models}

\def\baselinestretch{1}
\begin{table}
  \caption{Relative populations  for the states in the 3s$^2$ 3p$^2$ ground configuration
    of \ion{Fe}{xiii}.
}
\begin{center}
\begin{tabular}[c]{@{}rlclll@{}}
 \hline\hline \noalign{\smallskip}
 Model   & $^3$P$_{0}$ & $^3$P$_{1}$  & $^3$P$_{2}$ & $^1$D$_{2}$ &  $^1$S$_{0}$ \\
         &  10$^{-1}$   &  10$^{-1}$   & 10$^{-2}$  & 10$^{-3}$    & 10$^{-5}$ \\
\noalign{\smallskip}\hline\noalign{\smallskip}                                  
3065 states & 7.59 & 1.57  & 8.19 & 1.90  & 3.41 \\
749 states  & 7.60 & 1.57  & 8.13  & 1.88  & 3.37  \\
749 (no PE)  & 8.36 & 0.89  & 7.36  & 1.71  & 3.43  \\
100 states  & 7.64 & 1.55 & 7.91 & 1.79 & 3.26  \\
55 states  & 7.70  & 1.52  & 7.61  & 1.67  & 3.20 \\
sel. 55  states  & 7.67 & 1.54 & 7.75  & 1.74  & 3.25  \\
merged 33       & 7.63 & 1.55  & 7.91  & 1.81  & 3.29 \\
27  states  & 7.75 & 1.54  & 7.73  & 1.58  & 3.19  \\
sel. 55 states & & & & & \\
(P-CORONA) & 7.67 & 1.54  & 7.74 & 1.73  & 3.25\\
\noalign{\smallskip}\hline\noalign{\smallskip}                                  
\multicolumn{6}{l}{\small The models are at a distance of 1.1\rsun, at 1.4 MK, PE with } \\
\multicolumn{6}{l}{\small  a 6100 K black-body, and N$_{\rm e}$ = 10$^8$ cm$^{-3}$.} \\
\end{tabular}
\normalsize
\end{center}
\normalsize
\label{tab:pop1}
\end{table}

\begin{figure}
\centerline{\epsfig{file=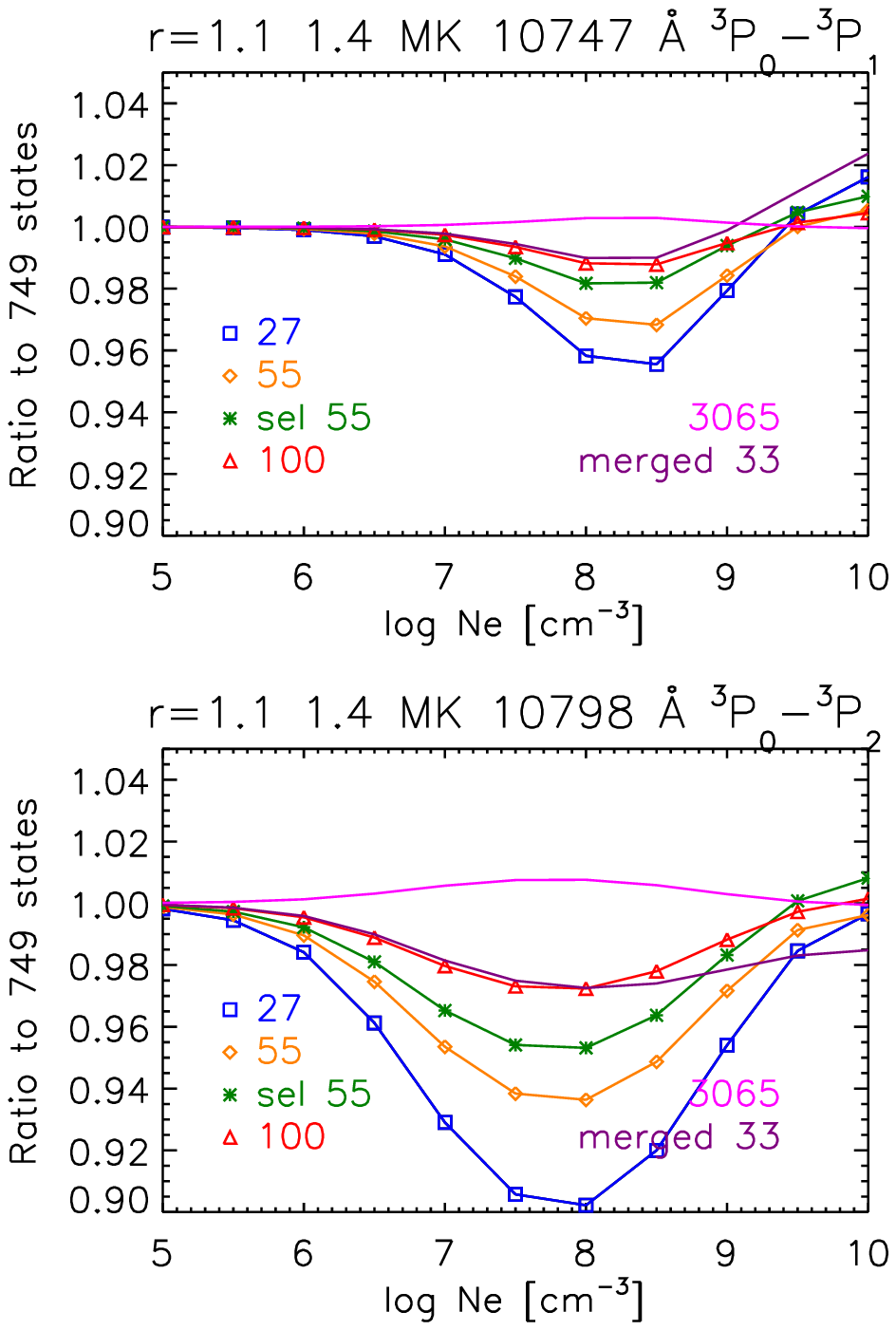, width=7.cm,angle=0 }}
\caption{The two plots show the ratio of the local emissivities in the
  two NIR lines obtained with 27, 91, and 200 states, relative to the values
  obtained with the current 749-state CHIANTI model.
  They also show the selected 55 states and the merged 31 states.
  The emissivities were calculated at a distance of 1.1\rsun, at 1.4 MK,
  PE with a 6100 K black-body, and as a function of electron density.   
}
\label{fig:ratios_1p1_1p4mk}
\end{figure}

\begin{figure}
\centerline{\epsfig{file=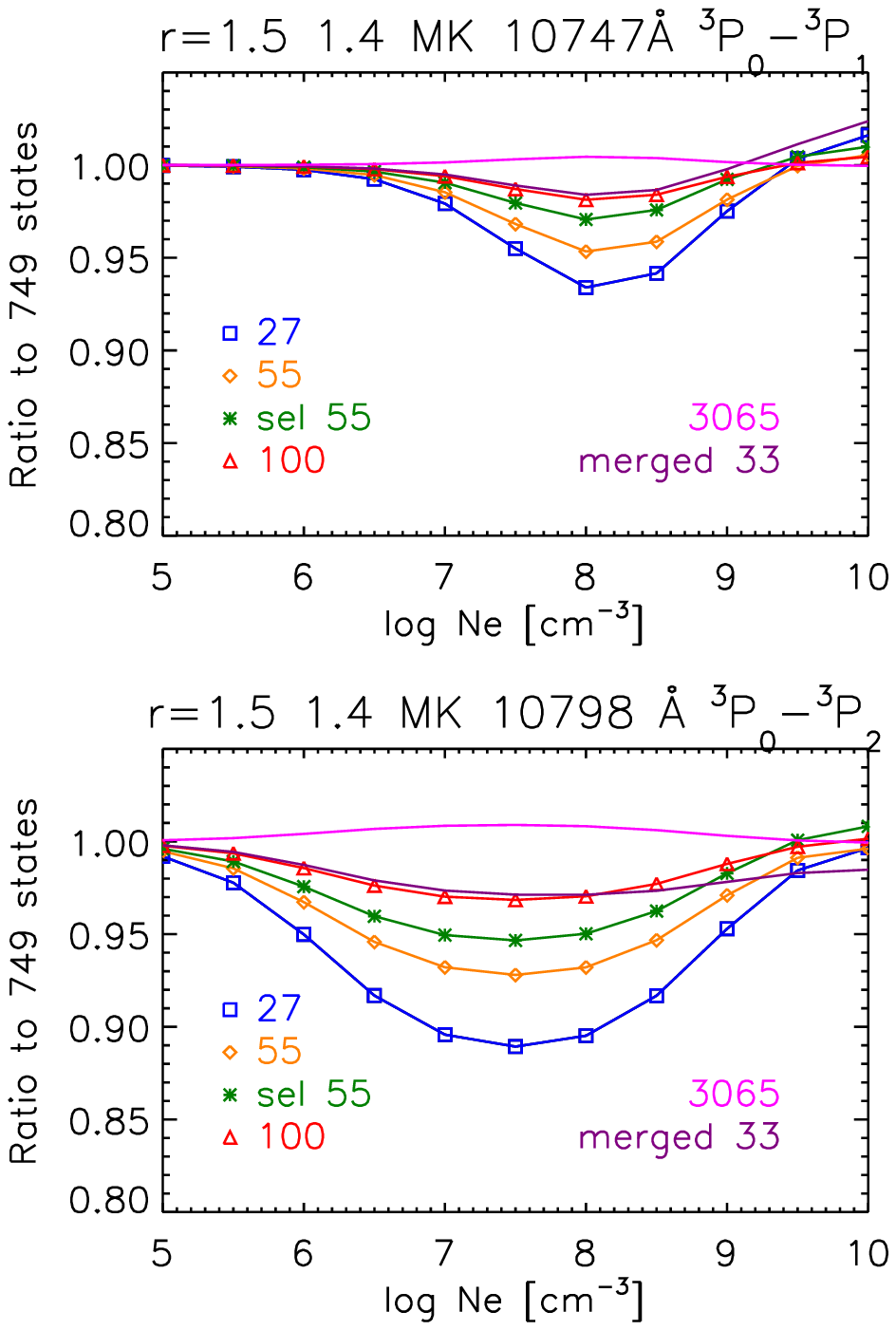, width=7.cm,angle=0 }}
\caption{The same as Figure~\ref{fig:ratios_1p1_1p4mk},
  calculated at a distance of 1.5\rsun\ at 1.4 MK.
}
\label{fig:ratios_1p5_1p4mk}
\end{figure}

We initially follow \cite{schad_dima:2020}, by plotting the ratio of the 
local emissivities in the
  two NIR lines obtained with a range of models, relative to the current 749-state CHIANTI model. The emissivities have been calculated with the CHIANTI IDL programs.
  As in \cite{schad_dima:2020} we have selected the lowest 27 and 100 states.
  In addition, we show in Figure~\ref{fig:ratios_1p1_1p4mk} the results of the 
  selected 55 states, as well as those of just selecting the lowest 55 states.
  Finally, we also show the ratios obtained with the more complete 3065 state
  model and with the reduced merged 33 state model. 
  Figure~\ref{fig:ratios_1p1_1p4mk} shows the results calculated for 
a distance of 1.1\rsun\ and an electron temperature of 1.4 MK, typical of the quiet Sun \citep[see, e.g.][]{gibson_etal:1999}.
  The results for the 27 and 100-states models appear to be 
  the same as those obtained by 
  \cite{schad_dima:2020}, as expected since they used the same atomic data. 
 Note that at such distance the electron density is typically  10$^8$ cm$^{-3}$.
  For the stronger line (measured by Lyot at 10746.80~\AA\ in air),
  all models are within 2\%,   except the 27-state model, where there is a discrepancy of about 5\%.
  For the weaker line (measured by Lyot at 10797.95~\AA\ in air), the 27-state model
  under-predicts the emissivity by about 10\%, while the selected 
  55-state model is within 5\%.
The reduced merged 33-state model performs extremely well, producing results 
close to the larger 100-state model, i.e. within about 2\% the 749-state model. 

\begin{table}
  \caption{Alignment for the states within the 
  3s$^2$ 3p$^2$ ground configuration
    of \ion{Fe}{xiii}, obtained  from P-CORONA and different atomic states}
\begin{center}
\begin{tabular}[c]{@{}rlclll@{}}
 \hline\hline \noalign{\smallskip}
 Model   &  $^3$P$_{1}$  & $^3$P$_{2}$ & $^1$D$_{2}$ & \\
  & 10$^{-2}$ & 10$^{-2}$ & 10$^{-3}$ & \\
\noalign{\smallskip}\hline\noalign{\smallskip}                                  
27 states  &  9.29  & 1.89  & -1.51  & \\

55 states  &  9.07 & 1.80 & -1.47 & \\

sel. 55 states & 8.91 & 1.77 & -1.43 & \\

100 states  & 8.77 & 1.72 & -1.38 & \\

200 states  & 8.74 & 1.71 & -1.37 & \\

749 states  & 8.58 & 1.66 & -1.30 & \\

\noalign{\smallskip}\hline\noalign{\smallskip}                                   
\multicolumn{5}{l}{\small The models are at a distance of 1.1\rsun, at 1.4 MK, PE with a } \\
\multicolumn{5}{l}{\small 6100 K black-body, and N$_{\rm e}$ = 10$^8$ cm$^{-3}$.} \\
\end{tabular}
\normalsize
\end{center}
\normalsize
\label{tab:align}
\end{table}

\begin{figure}
\centerline{\epsfig{file=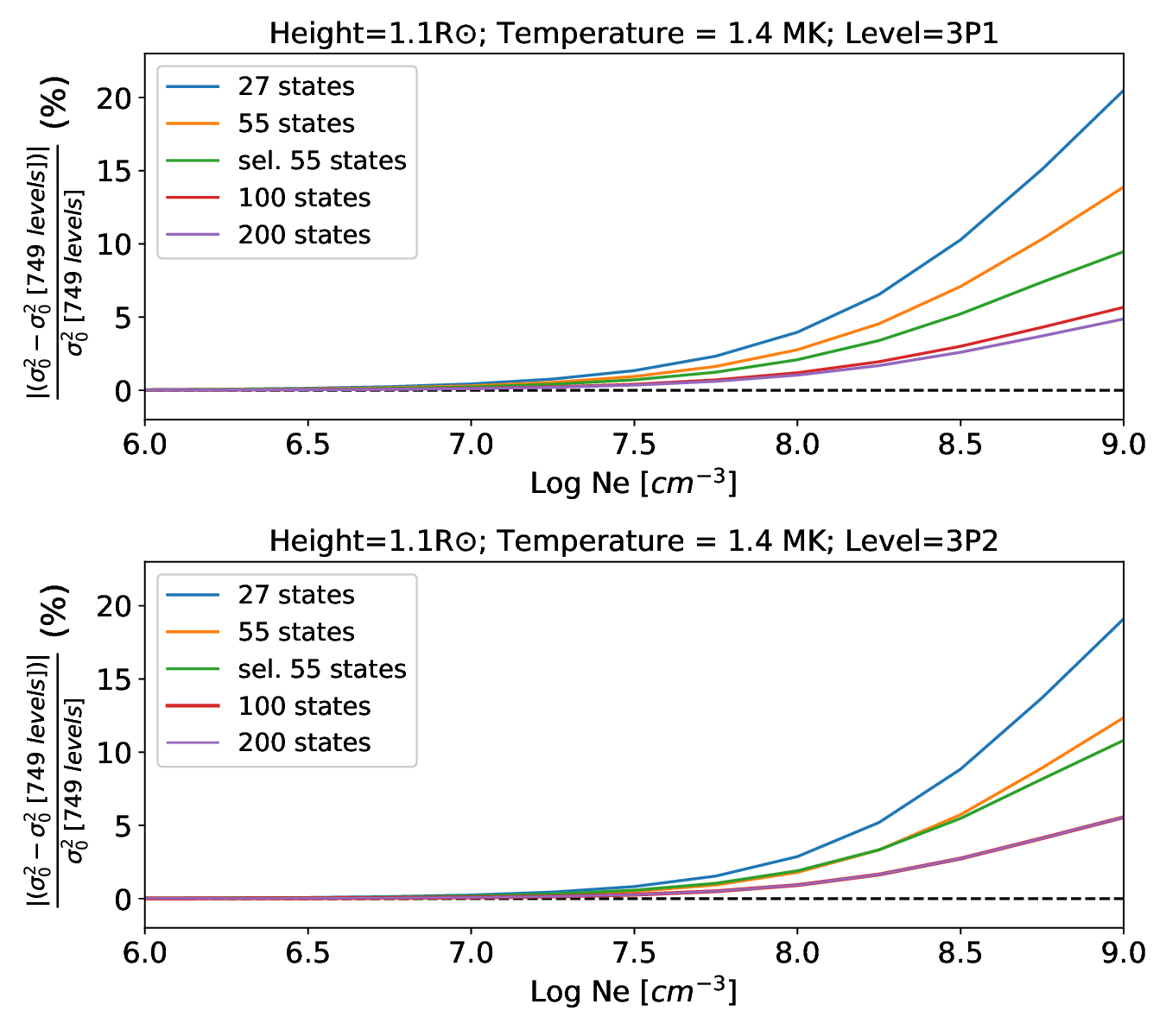, width=8.cm,angle=0 }}
\caption{The two panels show the percentage variation of the upper level alignment ($\sigma^2_0$) 
obtained with 27, 55 (both from CHIANTI 10 and reduced model), 100 and 200 states, relative to the values 
obtained with the 749-states for the Fe {\sc xiii} 10746.8\AA\ line. 
These computations were done for a distance of 1.1\rsun, at 1.4 MK,
PE with a 6100 K black-body, and as a function of electron density.
}
\label{fig:fe13alignment_1p1_1p4mk}
\end{figure}

\begin{figure}
\centerline{\epsfig{file=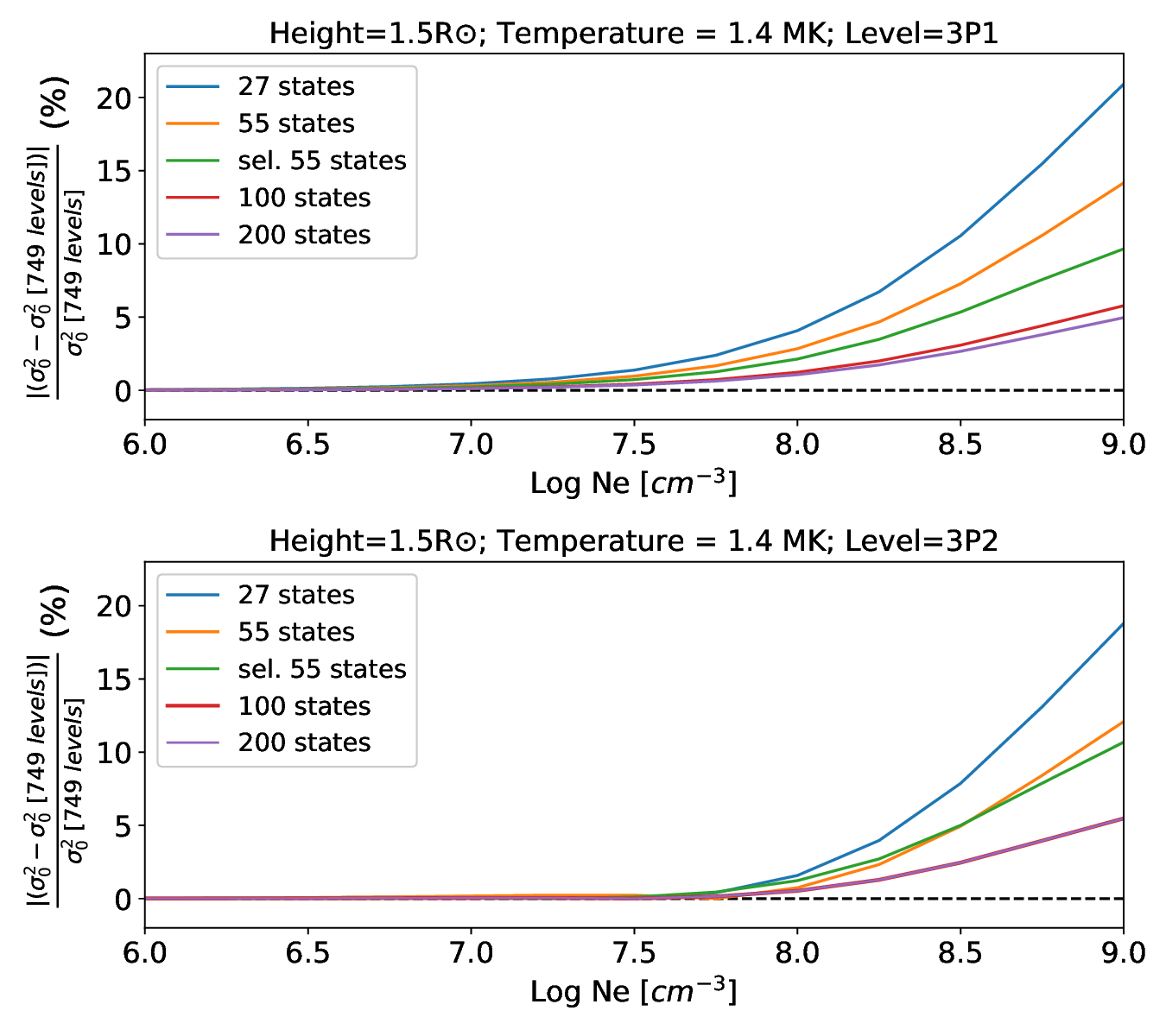, width=8.cm,angle=0 }}
\caption{The same as Figure~\ref{fig:fe13alignment_1p1_1p4mk} but at a distance of 1.5\rsun.
}
\label{fig:fe13alignment_1p5_1p4mk}
\end{figure}

\begin{figure*}
\centerline{\epsfig{file=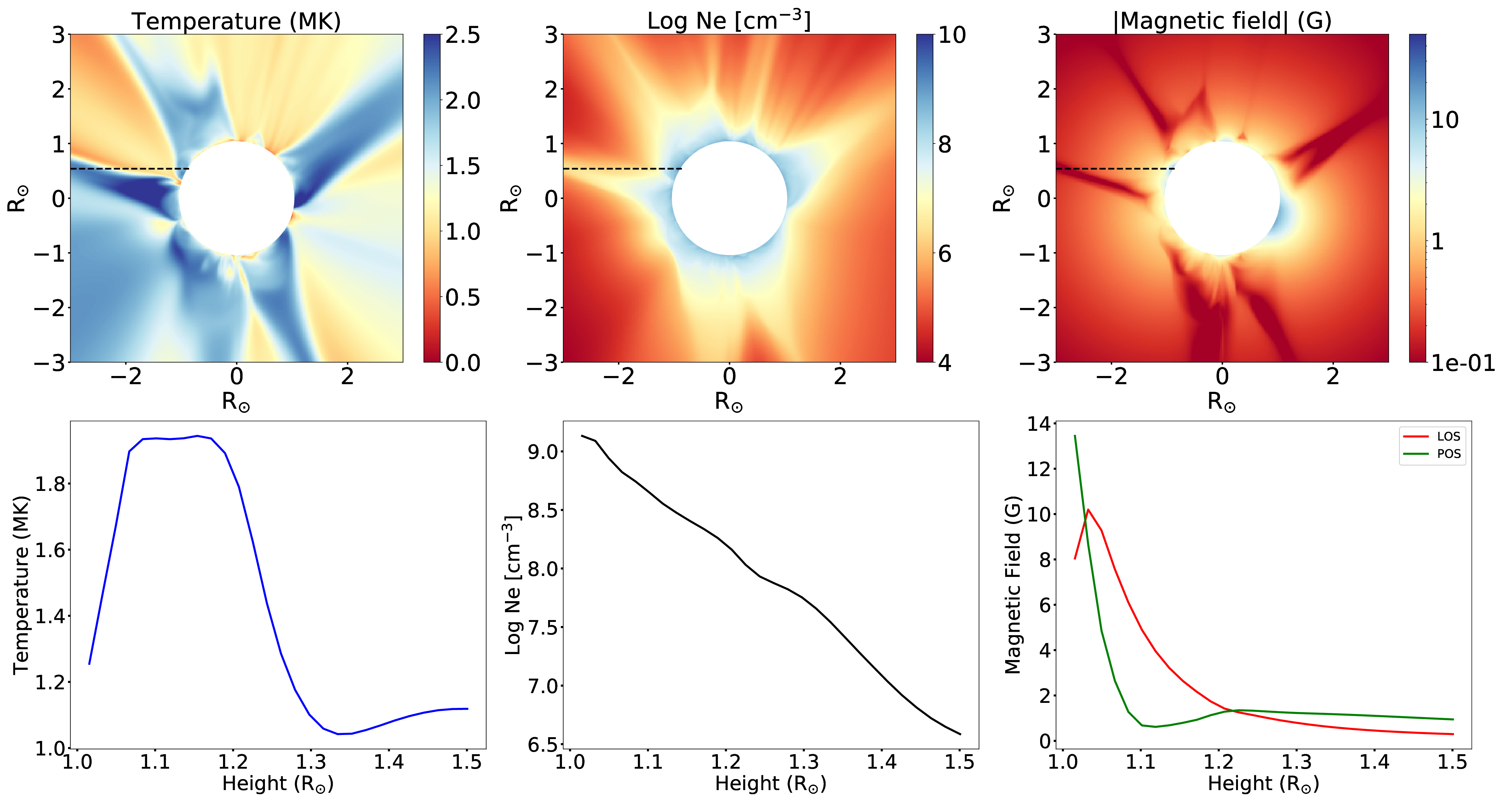, width=15.cm,angle=0 }}
\caption{The top panels show the temperature, electron number density, and magnetic field strength for the PSI Eclipse 2024 model in the plane of the sky (POS). The dashed line indicates the direction along which we consider the variation of these quantities, as shown in the bottom panel, used later in Figures~\ref{fig:fe13_10747stk} and \ref{fig:fe13_10798stk}. The third column in the bottom panel shows the plane-of-sky (POS) and line-of-sight (LOS) components of the magnetic field in the selected direction.}
\label{fig:eclipse2024PSI}
\end{figure*}

\begin{figure*}
\centerline{\epsfig{file=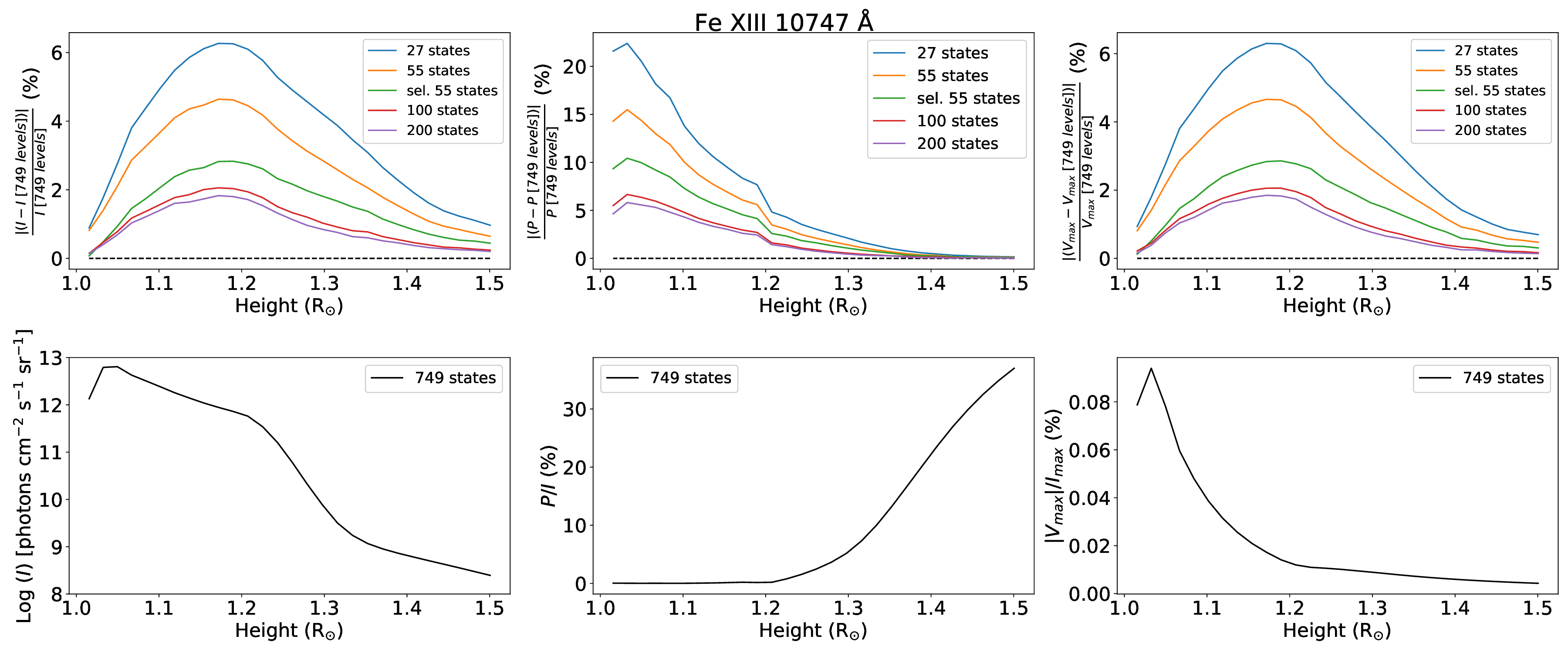, width=17.cm,angle=0 }}
\caption{The frequency integrated intensity, $I$, (in the first column) and polarization signals (the total linear polarization $P$ in the second column and the maximum of circular polarization $V_{max}$ in the last column) of the Fe {\sc xiii} 10747 \AA{} line. The spectro-polarimetric signals are computed in the PSI eclipse 2024 model along the dashed line indicated in Figure~\ref{fig:eclipse2024PSI}. The top panel shows the relative difference in the spectral signals and the bottom panel, the variation of them with height computed for the 749 states. }
\label{fig:fe13_10747stk}
\end{figure*}

\begin{figure*}
\centerline{\epsfig{file=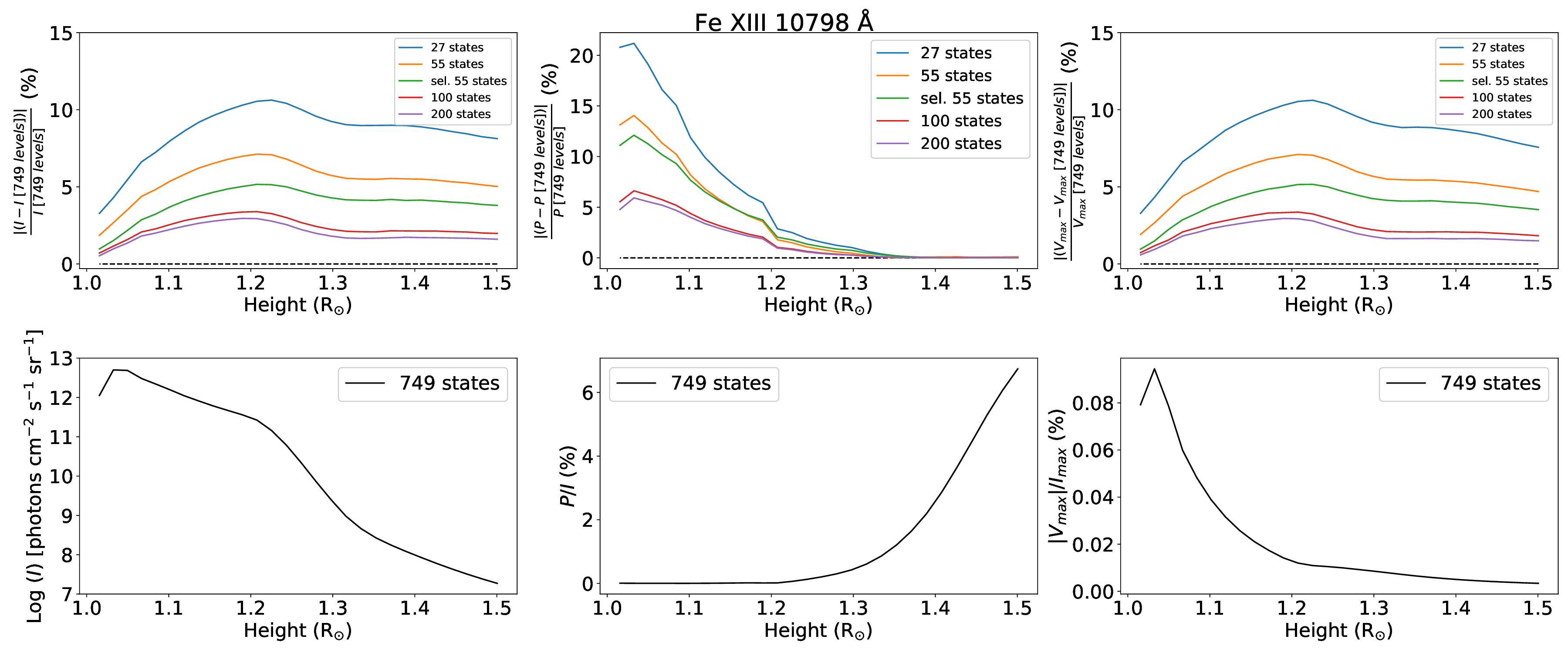, width=17.cm,angle=0 }}
\caption{Same as Figure~\ref{fig:fe13_10747stk} but for the Fe {\sc xiii} 10798 \AA{} line.
}
\label{fig:fe13_10798stk}
\end{figure*}

  The relative populations of the ground state are all reassuringly close
  to those obtained with the 749-state model, with the exception of the
  27-state model case, as shown in the Appendix.

  A similar picture is present  at a distance of 1.5\rsun,
  as shown in Figure~\ref{fig:ratios_1p5_1p4mk}. Note that in the quiet Sun
  at such distances one expects a density  less than 10$^7$ cm$^{-3}$,
  and as before the  selected 55-state model is within 5\% for the weaker line. 
  Similar plots at a slightly higher temperature of 1.8 MK adopted by
  \cite{schad_dima:2020} are shown in the Appendix in
  Figure~\ref{fig:ratios_1p5_1p8mk}.   The results are very similar.
When an active region is present, the ion emission is a superposition of the 
large-scale quiet Sun corona and a diffuse higher-temperature of about 2 MK.
Therefore, we expect the accuracy of the models to be similar also 
when modelling an active region.

Table~\ref{tab:pop1} shows the relative populations  for the main states in the 3s$^2$ 3p$^2$ ground configuration of \ion{Fe}{xiii}, 
for a distance of 1.1\rsun, at 1.4 MK,
 PE with a 6100 K black-body, and at an electron density of  10$^8$ cm$^{-3}$.
It is reassuring to see that the larger 3065 states model produces 
virtually the same populations as the CHIANTI 749 states model,
except the $^3$P$_{2}$ state, where the larger model produces a population 
0.7\% larger due to extra cascading. 

The third row in Table~\ref{tab:pop1} shows the populations
obtained without including PE, to confirm the large effect it has
even in the low corona. 
The following rows indicate the relative populations for the other models.
The last row shows the values obtained from \mbox{P-CORONA} using the selected 
55 states, to confirm that the CHIANTI and P-CORONA codes produce consistent 
results. 

Following this, using P-CORONA\footnote{https://polmag.gitlab.io/P-CORONA/index.html}, we compute the relative differences in the atomic alignment ($\sigma^2_0$) of the $^3$P$_{1}$ and $^3$P$_{2}$ levels in Figures~\ref{fig:fe13alignment_1p1_1p4mk} and \ref{fig:fe13alignment_1p5_1p4mk}. Atomic alignment quantifies the contribution to the linear polarization due to scattering with $\sigma^2_0 (J) = \frac{\rho^2_0 (J)}{\rho^0_0 (J)}$, where $\rho^K_Q$'s are the multipolar components of the atomic density matrix \citep[see][]{landi_book:2004}. Thus the atomic level alignment of the  $^3$P$_{1}$ and $^3$P$_{2}$ states contributes to the scattered polarization signal of the lines at 10747~\AA{} and 10798~\AA{} respectively. The relative difference in $\sigma^2_0$ is computed with respect to the 749 levels for the cases of 27, 55, selected 55, 100 and 200 states. 

The plasma conditions in Figures~\ref{fig:fe13alignment_1p1_1p4mk} and \ref{fig:fe13alignment_1p5_1p4mk} are the same as in Figures~\ref{fig:ratios_1p1_1p4mk} and \ref{fig:ratios_1p5_1p4mk} respectively. The relative difference in the atomic alignment in both the levels is below 5\% for coronal densities of 10$^8$ cm$^{-3}$ or below. The relative differences in atomic alignment increase as we consider higher densities. However, the overall atomic alignment remains low in these cases, as increased collisions tend to destroy the atomic alignment. Therefore, the large relative differences are a result of the small values involved. For a better comparison, we refer readers to Figure 4 in \citet{schad_dima:2020}, where the actual alignment values for the upper level of the relevant lines are plotted. 

Table~\ref{tab:align} shows the alignment for the states within the 
  3s$^2$ 3p$^2$ ground configuration
    of \ion{Fe}{xiii}, obtained  from P-CORONA and
    various reduced atomic models. 
    It is clear that the alignment is not much sensitive to the atomic models 
    for the representative plasma parameters considered.

\subsection{Computing times}

To show the importance of having a reduced atomic model when performing
SP large-scale calculations, we have run P-CORONA  for a single point in the 
corona, with typical parameters. 
The computing time, obtained with a single core, is shown in the second 
column of Table~\ref{tab:times}.  For a realistic 3D forward model, 
we have then scaled these times according to typical small volumes, 
e.g. a 257x512x643 MURaM simulation and the volume corresponding to a 
typical DKIST CryoNISRP  single-pointing slit scan, at 1\arcsec resolution.
In coronal mode, the 4' slit can be scanned by about 3', so we have 
considered a box of 240x240x240.

P-CORONA is parallelized (MPI+OpenMP), but even with a large number of cores
it is clear that reduced atomic models are really necessary. The computing time 
required in 100 states is 3.6 times larger than that of 55 states. 
The reduction in computational time achieved with the selected 55-state model, while maintaining  accuracy, is a significant advantage in addressing the complex problem of coronal line inversions. This efficiency allows for more feasible and timely analysis of coronal observations, particularly in large-scale SP studies.

\def\baselinestretch{1}
\begin{table}
  \caption{An estimate of the total computing time required for full forward modelling using P-CORONA 
  when different number of atomic levels are considered.}
\begin{center}
\begin{tabular}[c]{@{}rrrrr@{}}
 \hline\hline \noalign{\smallskip}
 No. of  levels   &  Single point  & MURaM model &  1\arcsec resolution \\ 
                  &                 &  (257x512x643) &  (240x240x240) \\
\noalign{\smallskip}\hline\noalign{\smallskip}                                  
27 states  &  2.3s &  54055 hrs&  8832 hrs& \\

55 states  &  7.4s &  173917 hrs& 28416 hrs& \\

100 states  &  26.9s&  632213 hrs& 103296 hrs& \\

749 states  & 438m &  12691276 hrs& 2073600 hrs&\\

\noalign{\smallskip}\hline\noalign{\smallskip}                                   
\end{tabular}
\normalsize
{For comparison, the computing times are calculated using a single core. P-CORONA is parallelized (MPI+OpenMP), so these numbers scale depending on the machine's efficiency and number of cores.}
\end{center}
\normalsize
\label{tab:times}
\end{table}

\subsection{Spectral synthesis with a realistic MHD model}

In this section, we show the results of the forward synthesis of the frequency-integrated intensity and polarization signals using 3D coronal MHD models by PSI. 
Our aim is to compute the differences in these signals in a realistic 3D MHD model 
while considering reduced atomic models. We took the recent PSI model corresponding to the total solar eclipse on 8 April, 2024{{\footnote{https://www.predsci.com/eclipse2024}}}.  
This was constructed using a time-dependent MHD model that was updated in near real-time with the latest measurements of the photospheric magnetic field {\citep{mikic_etal:2018,boe_etal:2021,boe_etal:2022,lionello_etal:2023}}. We use a time snapshot of this model, described in \cite{downs24}. The basic plasma parameters required for our computations, temperature, electron number density, and magnetic fields, are shown in Figure~\ref{fig:eclipse2024PSI}. We consider a 1D cut 
(indicated as dashed lines) in the plane of the sky (POS) and show in the bottom panels of Figure~\ref{fig:eclipse2024PSI} the variations of these parameters with height. 
In the bottom right panel of this Figure, we show both the POS and LOS magnetic fields in the chosen direction.
We have chosen such 1D cut for our representative computations with P-CORONA,
as a full 3D one would take too long when considering the larger-scale models.

We restrict our computations till 1.5 \rsun, beyond which the density drops below 10$^7$ cm$^{-3}$ and there would be very little signal
(also, DKIST CryoNIRSP can at most observe such distance). 
Our theoretical studies in the previous section show that the differences in the population and alignment of the concerned atomic levels does not change significantly with the number of atomic states used for these lower densities.   
We also find similar results, as shown in  Figure~\ref{fig:fe13_10747stk}
and Figure~\ref{fig:fe13_10798stk} for the Fe {\sc xiii} 10747 \AA{} and 10798 \AA{} lines. 

The bottom panels of these figures show 
the frequency integrated radiances $I$, the percentage of total linear polarization $P=\sqrt{Q^2+U^22}$ relative to $I$, and the percentage of maximum circular polarization $V_{max}$ relative to $I_{max}$. 
These values are calculated by integrating ALOS from -0.02 $\rsun$ to +0.02 $\rsun$, approximating the plane-of-sky values.
The upper panels in Figures~\ref{fig:fe13_10747stk}
and \ref{fig:fe13_10798stk} show the percentage differences between the 
results obtained with the various reduced models and the larger-scale 
749-state model. The 55-state {\it selected} model performs very well, with less
than 2\% differences. The relative difference in the total linear polarization is high close to the limb where the $P/I$ is too low to be easily measured.
Figure~\ref{fig:fe13stk-ratios} shows the relative differences in the Stokes values presented in the top panel of the Figures~\ref{fig:fe13_10747stk} and \ref{fig:fe13_10798stk}, but for the ratios $U/Q$ and $|V_{max}|/I_{max}$. The contribution from the higher levels through collisions and cascading almost cancel out in these ratios while considering the 27-state model or larger. However, this is not true for smaller atomic models as shown in Figure~\ref{fig:fe13stk-ratios}.

\section{Conclusions}

We have provided as examples two reduced atomic models 
for use in large-scale computations of  the intensities or the Stokes parameters of the  Fe {\sc xiii} 
NIR lines. These models provide significant savings in computing times.
Obviously, an alternative  way if one is interested in just the intensities
  would be to pre-calculate a lookup table, i.e. emissivities over a large parameter space
  in density, temperature and distance from the Sun, and then interpolate. 
We have shown with a range of calculations that line intensities and 
polarization signatures are still accurate to within 5\%, which is
comparable to the current uncertainties in the large-scale atomic model.

We demonstrated that the 55-state {\it selected} model performs well across a range of coronal conditions, preserving the essential physics of atomic transitions without the need for larger, more computationally demanding models. We also highlighted the efficiency of merging atomic states, which is especially useful for intensity calculations. While this method currently has limitations for polarization calculations, it still represents a highly accurate and computationally efficient approach. Other reduced models
can be provided upon request. 

The complexities of the optically thin solar corona are such that 
a better understanding of its physical state can only be achieved with 
routine large-scale forward models, to be compared to observations. 
One long-standing problem in solar physics has been the estimate 
of the local magnetic field. Recently, DKIST has shown that it is indeed capable of measuring the Stokes V, producing unprecedented polarized spectra of the  10747~\AA\ line  \citep{schad_etal:2024}. 
A combination of the unprecedented DKIST CryoNIRSP observations with 
large-scale computations with P-CORONA using reduced atomic models 
will finally provide the much needed information.  Though the Stokes ratios $U/Q$ and $V/I^{\prime}$, which contain the magnetic field information, are not significantly sensitive to cascading effects while considering large atomic models, the forward modelling computations required to calculate the individual Stokes parameters ($I$, $Q$, $U$, $V$) necessitate the use of reduced atomic models. Additionally, the forward modelling process, which forms the basis of inversion codes, can benefit from faster computations enabled by these reduced models.

Moving forward, the methodology used in developing reduced atomic models here will be extended to other spectral lines and ions, allowing their application in a wider range of coronal studies. This capability will be essential for future large-scale spectropolarimetric surveys, where computational efficiency is crucial for analyzing the vast amounts of data that will be collected and also in designing inversion methods. 

In summary, the reduced atomic models presented in this paper provide a practical and efficient solution for spectropolarimetric modelling in the solar corona, offering a balance between computational speed and accuracy.

\section*{Acknowledgments}
We are grateful to Cooper Downs (Predictive Science Inc.) for providing the PSI simulation.
GDZ acknowledges support from STFC (UK) via the consolidated grant
to the atomic astrophysics group  at DAMTP, University of Cambridge (ST/T000481/1).
This research was supported by the International Space Science Institute (ISSI) in Bern, through the ISSI International Team project \# 23-572 on  {\it Models and Observations of the Middle Corona}, led by GDZ.
HDS acknowledges support from the Agencia Estatal
de Investigación del Ministerio de Ciencia, Innovación
y Universidades (MCIU/AEI) under the grant “Polarimetric Inference of Magnetic Fields” and the European Regional Development Fund (ERDF) with reference PID2022-136563NB-I00/10.13039/501100011033.
HDS thanks Ángel de Vicente (IAC) for useful discussions. 
Finally, we thank the reviewers for useful comments.

\section*{Data Availability}
The CHIANTI-format models developed for this paper and 
the equivalent one in P-CORONA format are 
available on ZENODO at 
https://doi.org/10.5281/zenodo.13988111

Further data and programs will be posted in 
CHIANTI-VIP\footnote{www.chianti-vip.com}, a portal for advanced 
atomic data and models.

\bibliographystyle{mn2e}

\bibliography{fe_13r}

\clearpage

 \appendix

\section{Extra material}

\begin{figure}
\centerline{\epsfig{file=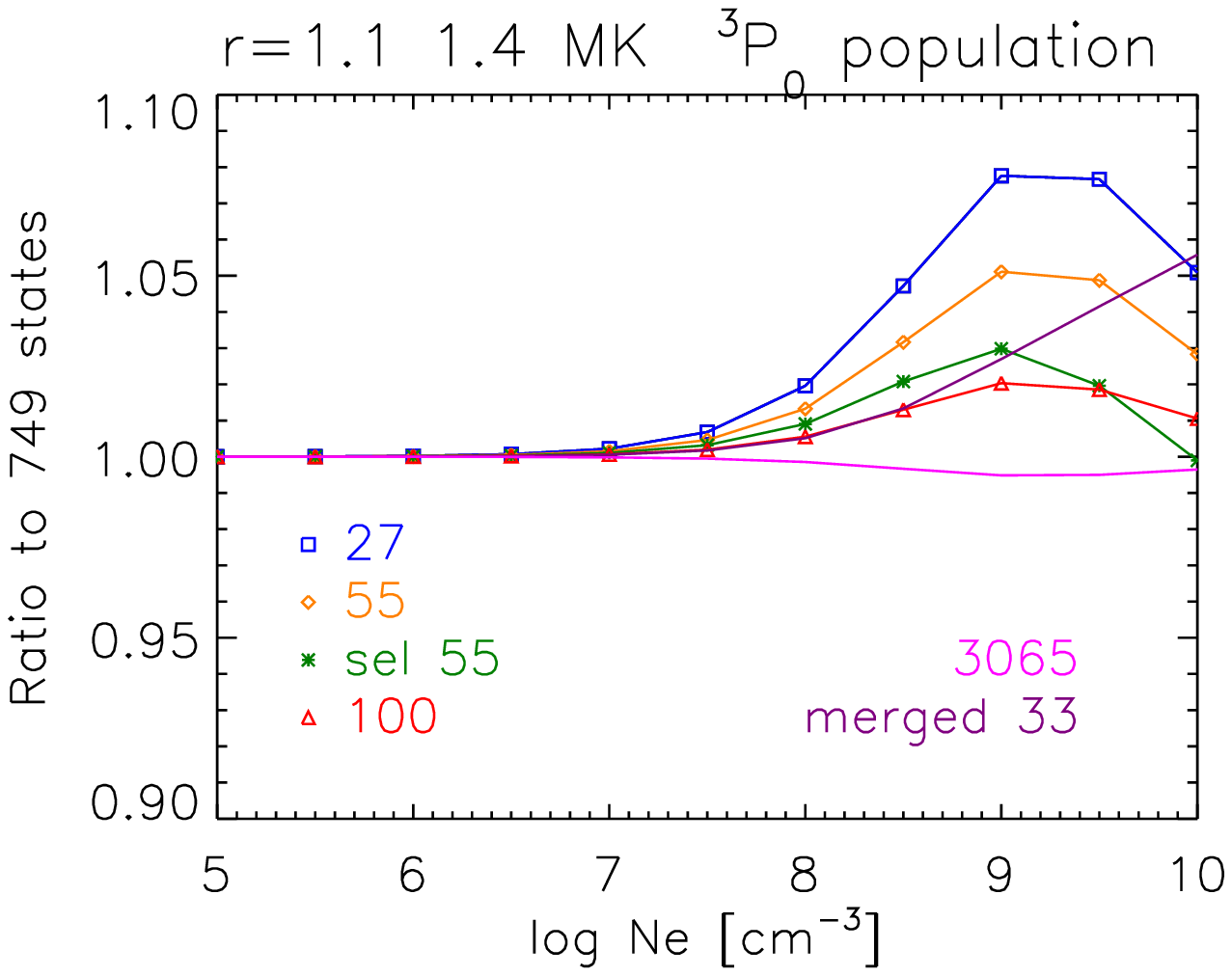,width=6.cm,angle=-90 }}
\centerline{\epsfig{file=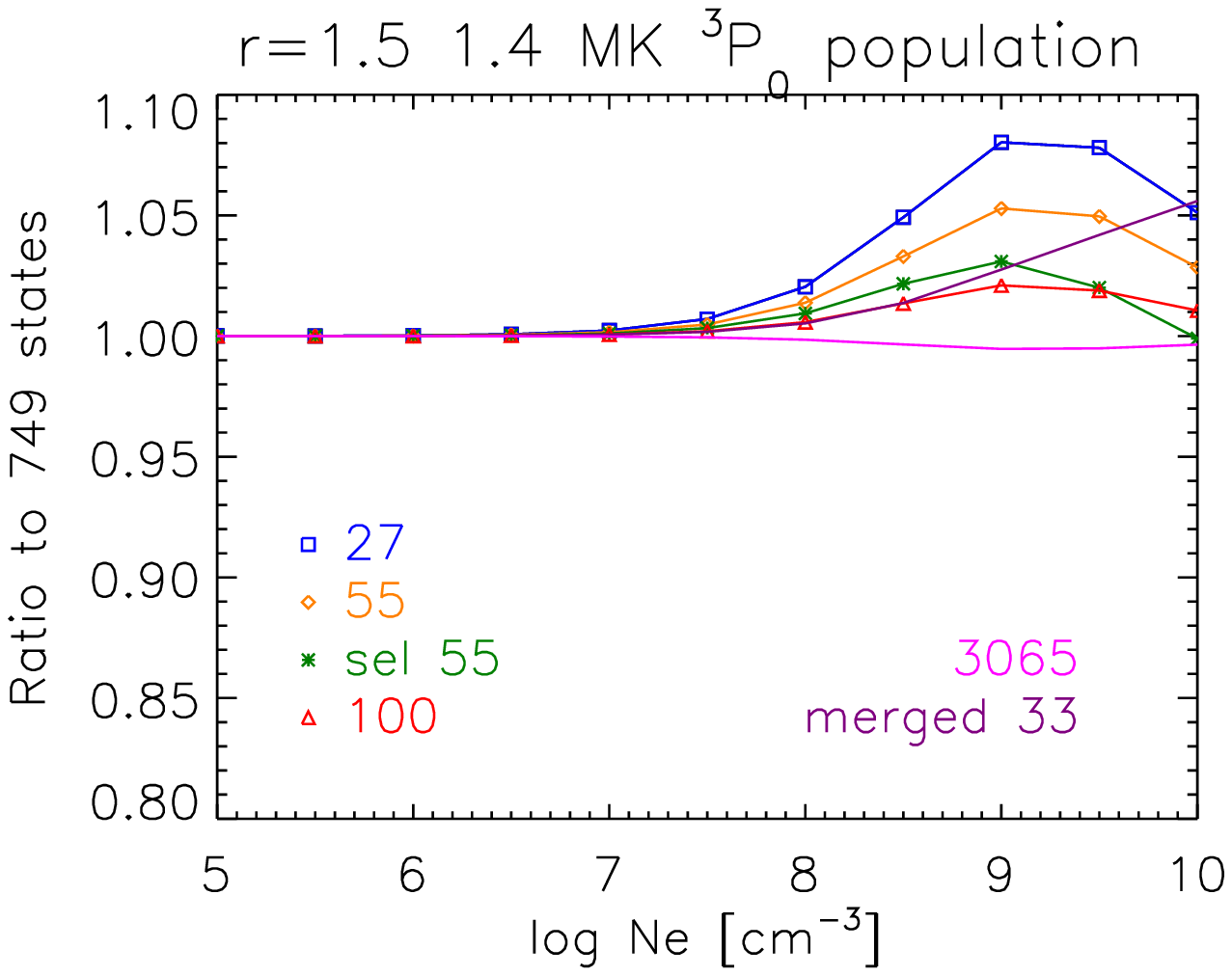,width=6.cm,angle=-90 }}
\centerline{\epsfig{file=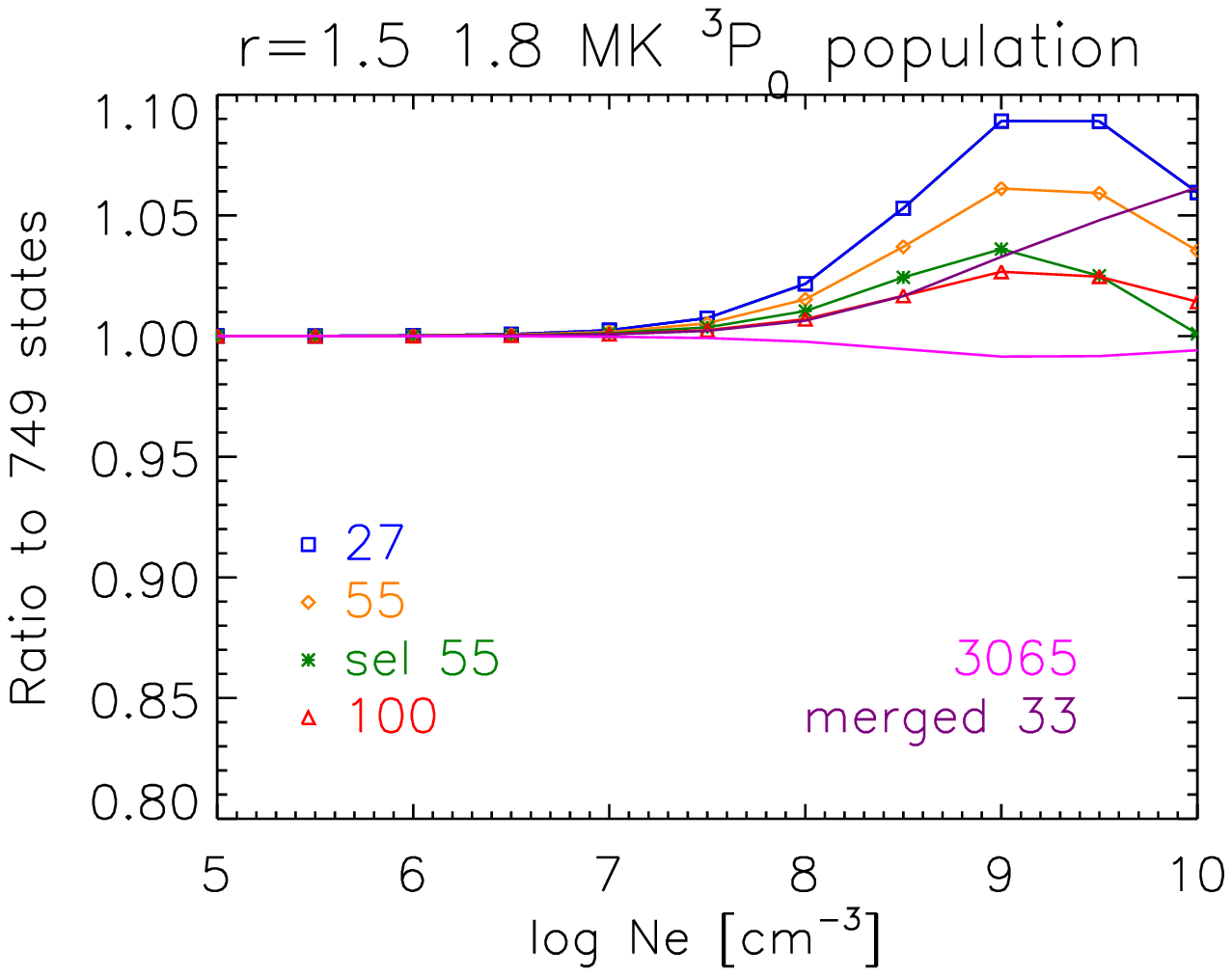,width=6.cm,angle=-90 }}
\caption{The populations of the ground state, 
relative to the values  obtained with the 749-states model, for the temperatures and 
distances listed.
}
\label{fig:ratios_1p1_1p4mk_gs}
\end{figure}

Figure~\ref{fig:ratios_1p1_1p4mk_gs} shows the populations of the ground state, 
relative to the values  obtained with the 749-states model, for the two sets of 
parameters chosen for the main paper and for the case presented by \cite{schad_dima:2020},
a distance of 1.5\rsun\ and a temperature of 1.8 MK.
It is clear that differences are negligible, except in cases where the local 
density would be very high as in the core of an active region. In such a case
the excited states become populated more and the differences reach a 
few percent. 

\begin{figure}
\centerline{\epsfig{file=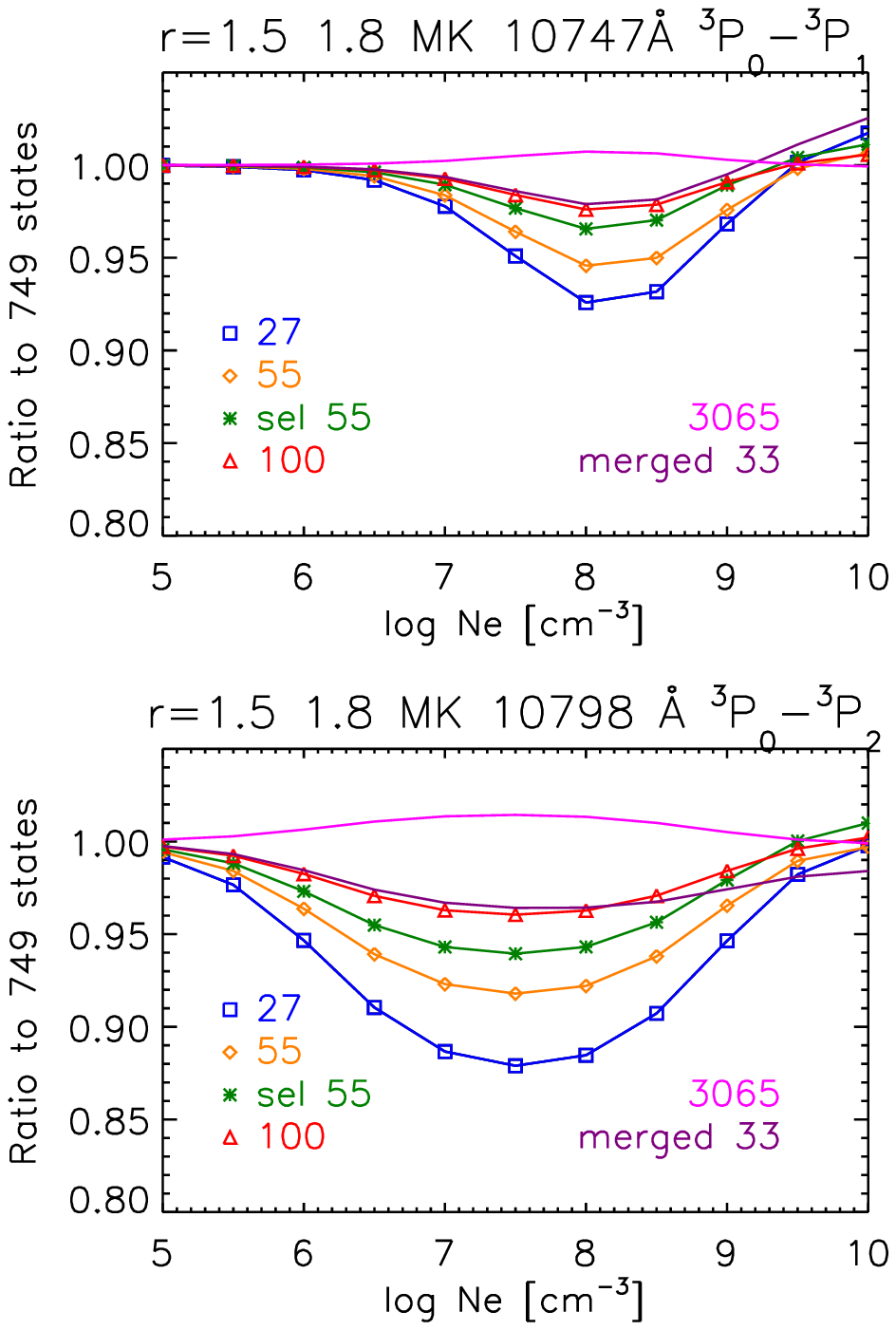, width=7.cm,angle=0 }}
\caption{The same as Figure~\ref{fig:ratios_1p1_1p4mk},
  calculated at a distance of 1.5\rsun\ at 1.8 MK.
}
\label{fig:ratios_1p5_1p8mk}
\end{figure}

\begin{figure}
\centerline{\epsfig{file=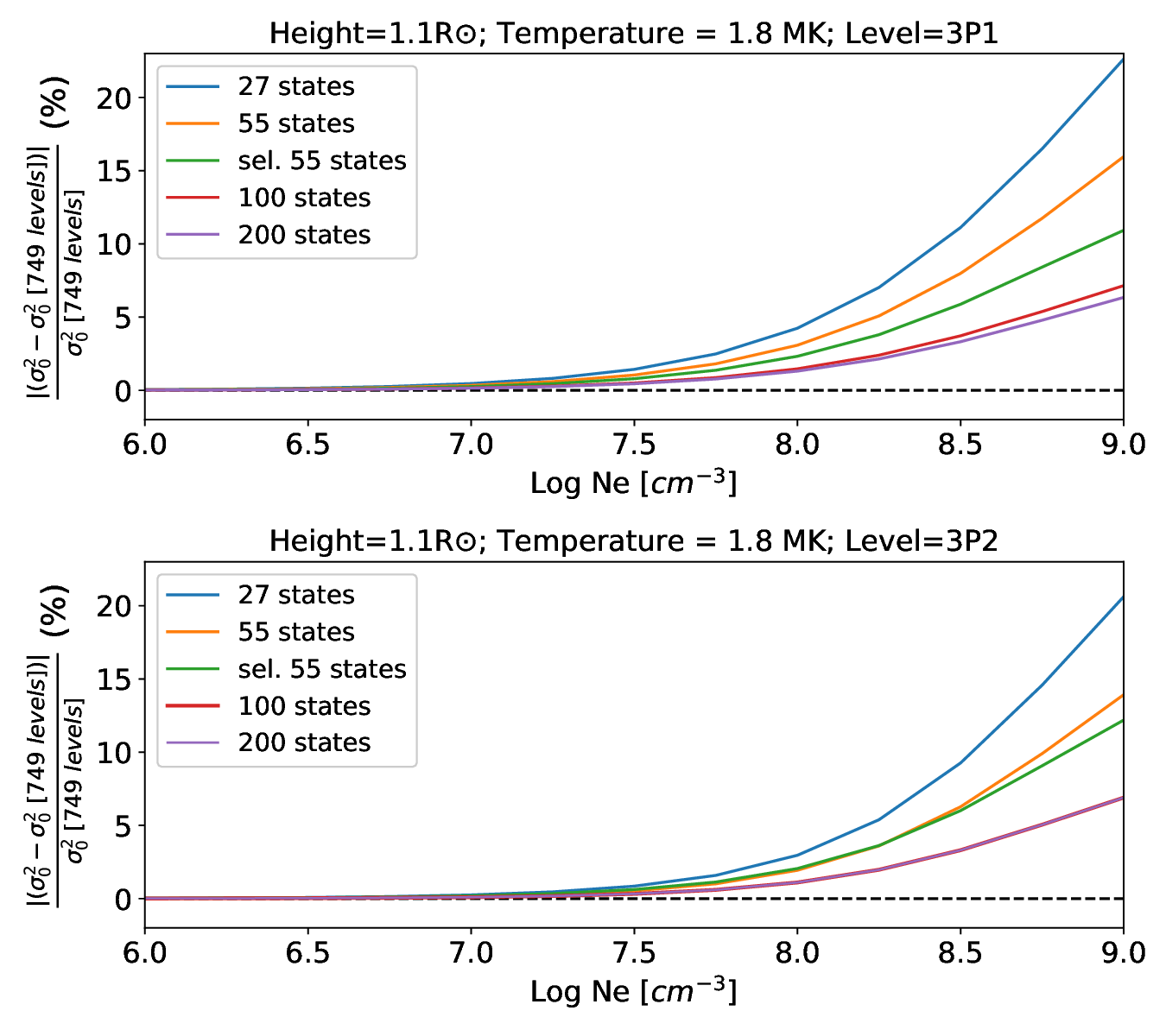, width=8.cm,angle=0 }}
\caption{The same as Figure~\ref{fig:fe13alignment_1p1_1p4mk} but for a temperature of 1.8 MK.
}
\label{fig:fe13alignment_1p1_1p8mk}
\end{figure}

\begin{figure}
\centerline{\epsfig{file=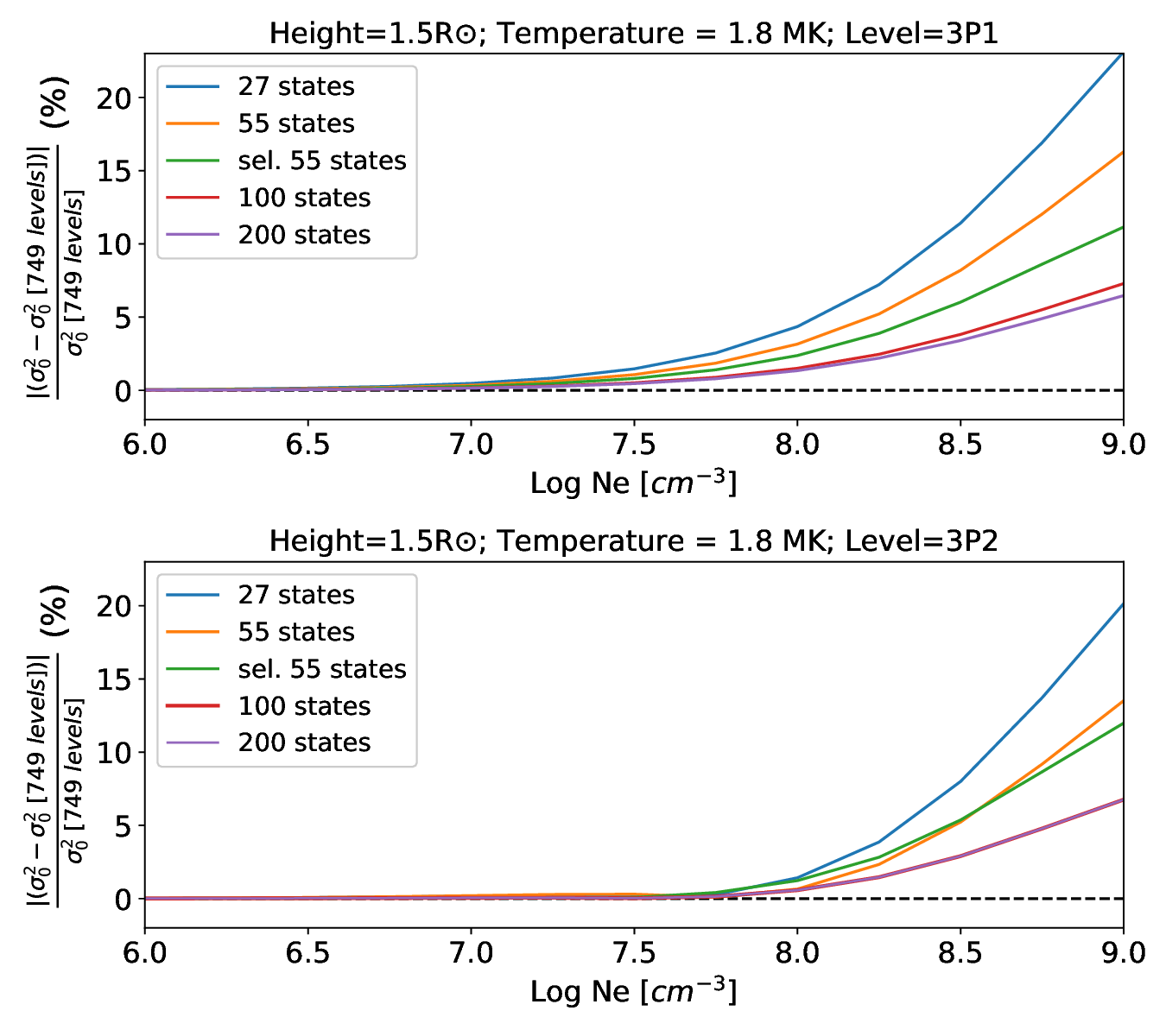, width=8.cm,angle=0 }}
\caption{The same as Figure~\ref{fig:fe13alignment_1p5_1p4mk} but for a temperature of 1.8 MK.
}
\label{fig:fe13alignment_1p5_1p8mk}
\end{figure}

Figure~\ref{fig:ratios_1p5_1p8mk} shows the relative intensities of the two NIR lines,
as in Figure~\ref{fig:ratios_1p1_1p4mk},
  calculated at a distance of 1.5\rsun\ at 1.8 MK, for comparison with the results 
  shown by  \cite{schad_dima:2020}. Figures~\ref{fig:fe13alignment_1p1_1p8mk} and \ref{fig:fe13alignment_1p5_1p8mk} show the variation of alignment for the above mentioned cases.

\begin{figure*}
\centerline{\epsfig{file=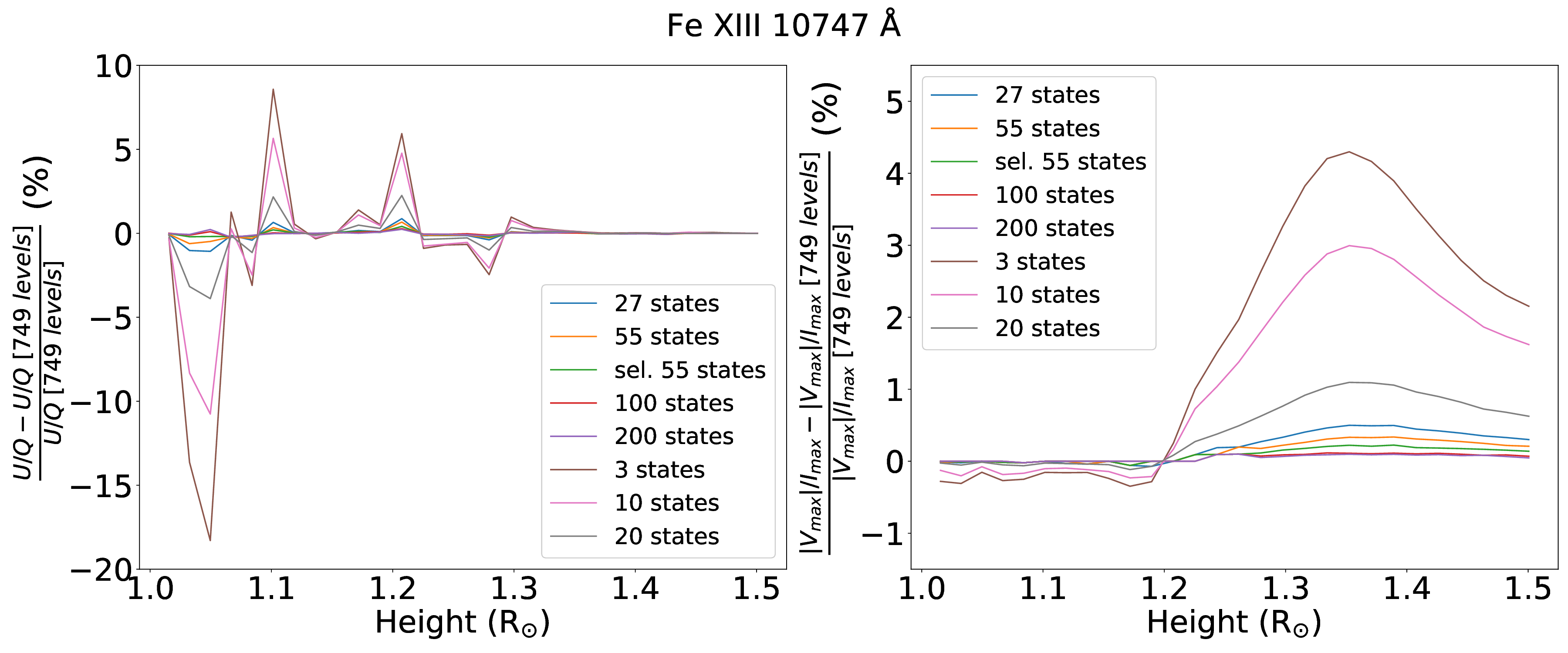, width=12.cm,angle=0 }}
\centerline{\epsfig{file=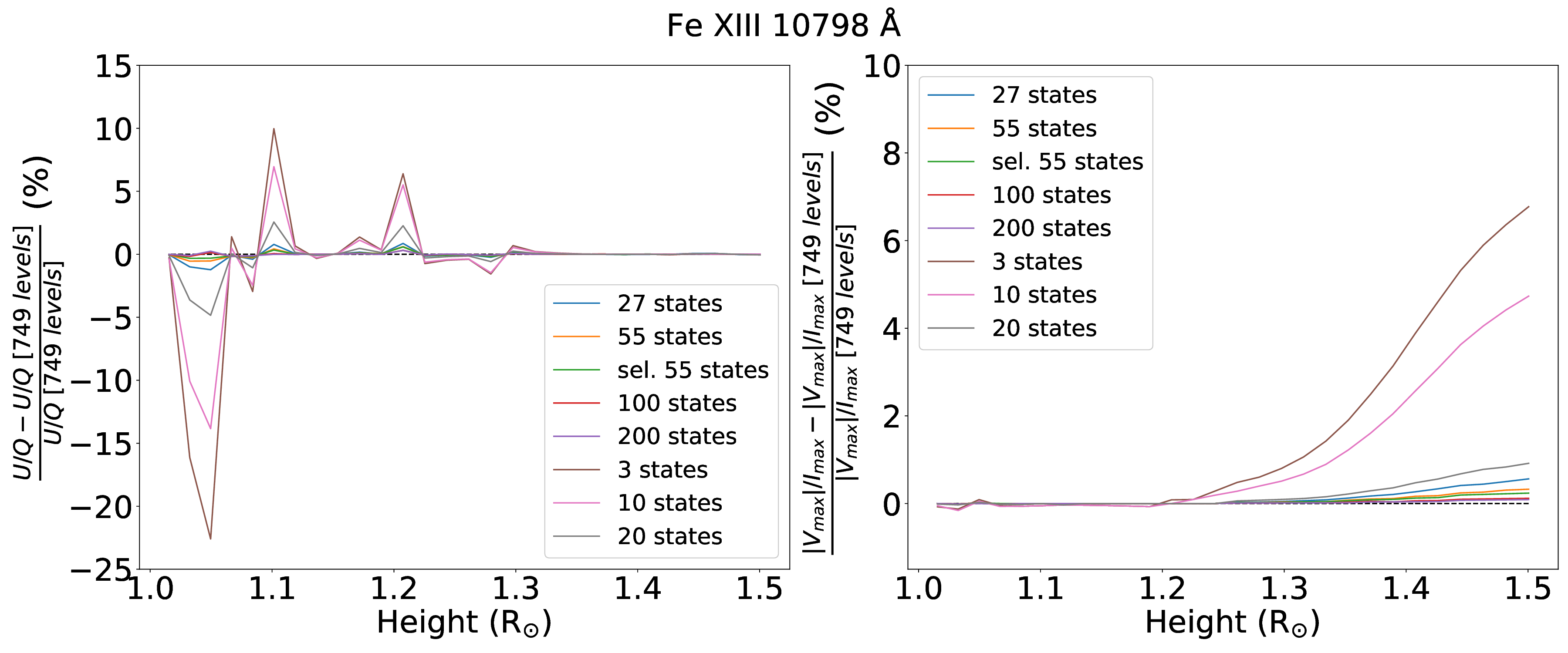, width=12.cm,angle=0 }}
\caption{The left and right panels show the relative differences in the ratios $U/Q$ and $|V_{max}|/I_{max}$, respectively, for 3, 10, 20, 27, 55, 100, and 200 states compared to the 749 levels computed in the PSI eclipse 2024 model, as shown in the bottom panel of Figure \ref{fig:eclipse2024PSI}. The top and bottom panels correspond to the Fe {\sc xiii} 10747 \AA{} and 10798 \AA{} lines, respectively. 
}
\label{fig:fe13stk-ratios}
\end{figure*}

 Figure~\ref{fig:fe13stk-ratios} shows the relative differences in $U/Q$ and $|V_{max}|/I_{max}$ for different atomic levels considered compared to the full 749 level calculations for both the Fe {\sc xiii} lines. We compute the simpler frequency integrated $|V_{max}|/I_{max}$ 
  as a proxy for $V/I^{\prime}$. These results are computed for the 1D variation in the PSI eclipse 2024 model shown in Figure \ref{fig:eclipse2024PSI}. Along with the atomic models containing 27, 55, 100, and 200 states, we additionally compute the relative differences for models with 3, 10, and 20 states to highlight how the relative differences are significantly larger when cascading and collisional effects from higher states are not considered. 

\end{document}